\documentclass[journal]{elsarticle}

\journal{Intelligent Systems with Applications}

\usepackage{lineno,hyperref}
\usepackage{graphicx}
\usepackage[justification=centering]{caption}
\usepackage{balance}
\usepackage{subfigure}
\usepackage{color}
\usepackage{supertabular}
\usepackage{longtable}

\begin{document}
\begin{frontmatter}

\title{Using 5G in Smart Cities: \\A Systematic Mapping Study}

\author{Chen Yang\textsuperscript{a,b,c}, Peng Liang\textsuperscript{b\textsuperscript{*\corref{cor1}\fnref{label1}}}, Liming Fu\textsuperscript{b}, Guorui Cui\textsuperscript{b}, Fei Huang\textsuperscript{b}, Feng Teng\textsuperscript{c}, Yawar Abbas Bangash\textsuperscript{d}}
\affiliation[a]{organization={School of Artificial Intelligence, Shenzhen Polytechnic}, city={Shenzhen}, country={China}}
\affiliation[b]{organization={School of Computer Science, Wuhan University}, city={Wuhan}, country={China}}
\affiliation[c]{organization={IBO Technology (Shenzhen) Co., Ltd.}, city={Shenzhen}, country={China}}
\affiliation[d]{organization={School of Computer Science, National University of Sciences and Technology}, city={Islamabad}, country={Pakistan}}
\fntext[label1]{Corresponding author at: School of Computer Science, Wuhan University, China. Tel.: +86 27 68776137; fax: +86 27 68776027. E-mail address: liangp@whu.edu.cn (P. Liang).}


\begin{abstract}
5G is the fifth generation wireless network, with a set of characteristics, such as high bandwidth and data rates, massive connectivity, broad coverage, and low latency. The scenarios of using 5G include enhanced Mobile Broadband (eMBB), massive Machine Type Communications (mMTC), and ultra-Reliable and Low-Latency Communications (uRLLC). 5G is expected to support a wide variety of applications, such as city management, healthcare, transportation, agriculture, and energy management. In this paper, we conducted a systematic mapping study that covers the literature published between January 2012 and December 2019 regarding using 5G in smart cities. The scenarios, architecture, technologies, challenges, and lessons learned of using 5G in smart cities are summarized and further analyzed based on 32 selected studies, and the results are that: (1) The studies are distributed over 27 publication venues. 17 studies report results based on academic studies and 13 studies use demonstration or toy examples. Only 2 studies report using 5G in smart cities based on industrial studies. 16 studies include assumptions of 5G network design or smart city scenarios.
(2) The most discussed smart city scenario is transportation, followed by public safety, healthcare, city tourism, entertainment, and education.
(3) 28 studies propose and/or discuss the architecture of 5G-enabled smart cities, containing smart city architecture (treating 5G as a component), 5G network architecture in smart cities, and business architecture of using 5G in smart cities.
(4) The most mentioned 5G-related technologies are radio access technologies, network slicing, and edge computing. 
(5) Challenges are mainly about complex context, challenging requirements, and network development of using 5G in smart cities.
(6) Most of the lessons learned identified are benefits regarding 5G itself or the proposed 5G-related methods in smart cities.
This work provides a reflection of the past eight years of the state of the art on using 5G in smart cities, which can benefit both researchers and practitioners in this field.
\end{abstract}


\end{frontmatter}

\section{Introduction}\label{section1}

Urbanization is developing rapidly worldwide and more than 54\% of the populations in the world have moved into cities \cite{xing2016thetrendes}. As an example, in China, including Beijing, Shanghai, Guangzhou, and Shenzhen, many cities have more than 10 million residents. Cities are growing larger and more complex in recent years, leading to many risks, concerns, and challenges in their management \cite{nam2011conceptualizing}. The limitation of traditional city management patterns becomes a critical issue. To overcome the problems caused by population, resources, environment, among others, there is a need to make cities smart. In this study, we adopted the definition from \cite{nam2011conceptualizing} and \cite{moss2009informed}: ``\textit{smart city is an organic connection among technological, human, and institutional components. A smart city should be treated as an organic whole, i.e., as a network as well as a linked system.}'' 

The ways in which people live in smart cities have significant changed due to various technological advancements. There are many towering skyscrapers in smart cities, replacing old buildings. They are much  smarter, not only because of the innovation of the science of building materials, but also due to the rapidly development of sensor technology and convergence of new data streams.
Among all the technologies, 5G is the fifth generation wireless network, which is considered as one of the most important components for future smart cities \cite{agiwal2016next}. 5G networks are going to be deployed or have already been deployed in many cities, which are underway to change our lives. As mentioned by Setyawan \textit{et al.}, the number of smartphone users will reach 5.9 billion in 2025 \cite{setyawan2020briefreview}. Arshad \textit{et al.} reported that the number of devices on Internet will gain over 100 billion in 2025 \cite{arshad2017greeniot}. In the 5G era, real-time monitoring and controlling in smart cities will be possible instead of only a buzzword, addressing the challenges of overpopulation and advancing urbanization of cities. 
5G includes a set of characteristics, such as high bandwidth and data rates, massive connectivity, broad coverage, and low latency \cite{agiwal2016next}. 5G aims to address three cases, i.e., enhanced Mobile Broadband (eMBB), massive Machine Type Communications (mMTC), and ultra-Reliable and Low-Latency Communications (uRLLC) \cite{morgado2018survey}. In the smart city context, the eMBB case covers the exchange of data between various user equipment or surveillance cameras and edge or cloud servers in the system. The data includes text, audio, and video, which may be characterized by large bandwidth requirements. The mMTC case covers a large number of connected devices (e.g., sensors and wearable devices) through a dense deployment in a city. These devices are used to provide different services (e.g., automatic monitoring of buildings). The uRLLC case covers those communications that are time-critical or require high delivery probability. An example is communications among driver-less cars, edge devices, and base stations in a city. 
Though 5G is believed a game changer in smart cities, various challenges and unknowns emerge in the three cases of using 5G in smart cities. Examples include how to address the power supply problem for a huge amount of devices, how 5G works together with existing 4G networks, how to manage wide distribution of the devices especially those devices in remote or inaccessible areas, and possible high costs of constructing and maintaining 5G networks. Such challenges and unknowns bring limitations for realizing 5G in smart cities, drive researchers and practitioners to develop new approaches, methods, techniques, and tools. Currently many cities are actively prompting the usage of 5G and researchers and practitioners have conducted various trials in those cities (e.g., \cite{antonelli2018city}\cite{nizzi2018evaluation}\cite{rusti20195g}\cite{Okumura}).

\subsection{Motivation and contribution of the study}

Many technologies are considered to be promising to improve the smartness of cities, including 5G, Artificial Intelligence (AI), Internet of Things (IoT), big data, and cloud computing \cite{law2019smart}. Unlike those technologies (e.g., AI, IoT, big data, and cloud computing) that have been evolved over many years, there is a lack of systematic analysis of the state of the art regarding using 5G in smart cities. Given the importance of using 5G in smart cities and the lack of systematic review on the topic, we designed and executed a systematic mapping study (SMS) on the topic of using 5G in smart cities. The objective of the study is to analyze the usage of 5G in smart cities for the purpose of exploration and analysis with respect to the scenarios, architecture, 5G-related technologies, challenges (problems), and lessons learned concerning using 5G in smart cities. In this work, our aim is not to propose a new approach, method, technique, or tool, but use a review method (i.e., SMS) to analyze the literature published between January 2012 and December 2019 regarding using 5G in smart cities.

Compared to literature surveys and systematic literature reviews (SLRs), SMSs do not aim at synthesizing evidence. Instead, SMSs are usually used to provide a wide but more coarse-grained overview of a research area, identify research evidence on a topic, and present classifications and quantitative results. As an example, instead of focusing on every specific challenge regarding using 5G in smart cities as well as their reasons and possible solutions, we used the Constant Comparison method \cite{glaser1968discovery} to generate incidents, concepts, and categories. The output is a classification of the challenges with examples. The process of performing an SMS is composed of the following steps: (1) study search: search studies regarding a specific topic from websites, such as the ACM Digital Library, (2) study selection: identify relevant studies based on a set of criteria through their title, abstract, and full text from the search results, (3) data extraction: extract original words, sentences, or paragraphs from the selected studies, and (4) data analysis: analyze the extracted data to answer the research questions (RQs). Before designing and conducting any further analysis (e.g., through an SLR), this SMS is the first step to provide a reflection of the past eight years of research and practice on using 5G in smart cities, which can benefit both researchers and practitioners in this field. Through classifying the studies to identify clusters of the studies on five aspects: scenario, architecture, technology, challenge (i.e., gaps indicating the need for more primary studies), and lesson learned, we structured the research area of using 5G in smart cities and formed the basis of an SLR with more synthesis regarding this topic as future work.

In this paper, ``\textit{scenarios}'' means smart city scenarios (e.g., transportation or public safety in a city) that include various smart city components (e.g., devices) for city and its infrastructure management; ``\textit{architecture}'' refers to both technical architecture and business architecture of using 5G in smart cities; ``\textit{5G technologies}'' means the technologies that can be used in 5G network, such as millimeter wave (mm-wave), beamforming, and radio access; ``\textit{challenges}'' refers to the problems of using 5G in smart cities; ``\textit{lessons learned}'' is regarding the experience of using 5G in smart cities based on evaluations (e.g., experiments and case studies), instead of experience about how to conduct evaluations.

\subsection{Structure of the study}
\begin{figure*}[!t]
\centering
\includegraphics[width=5in]{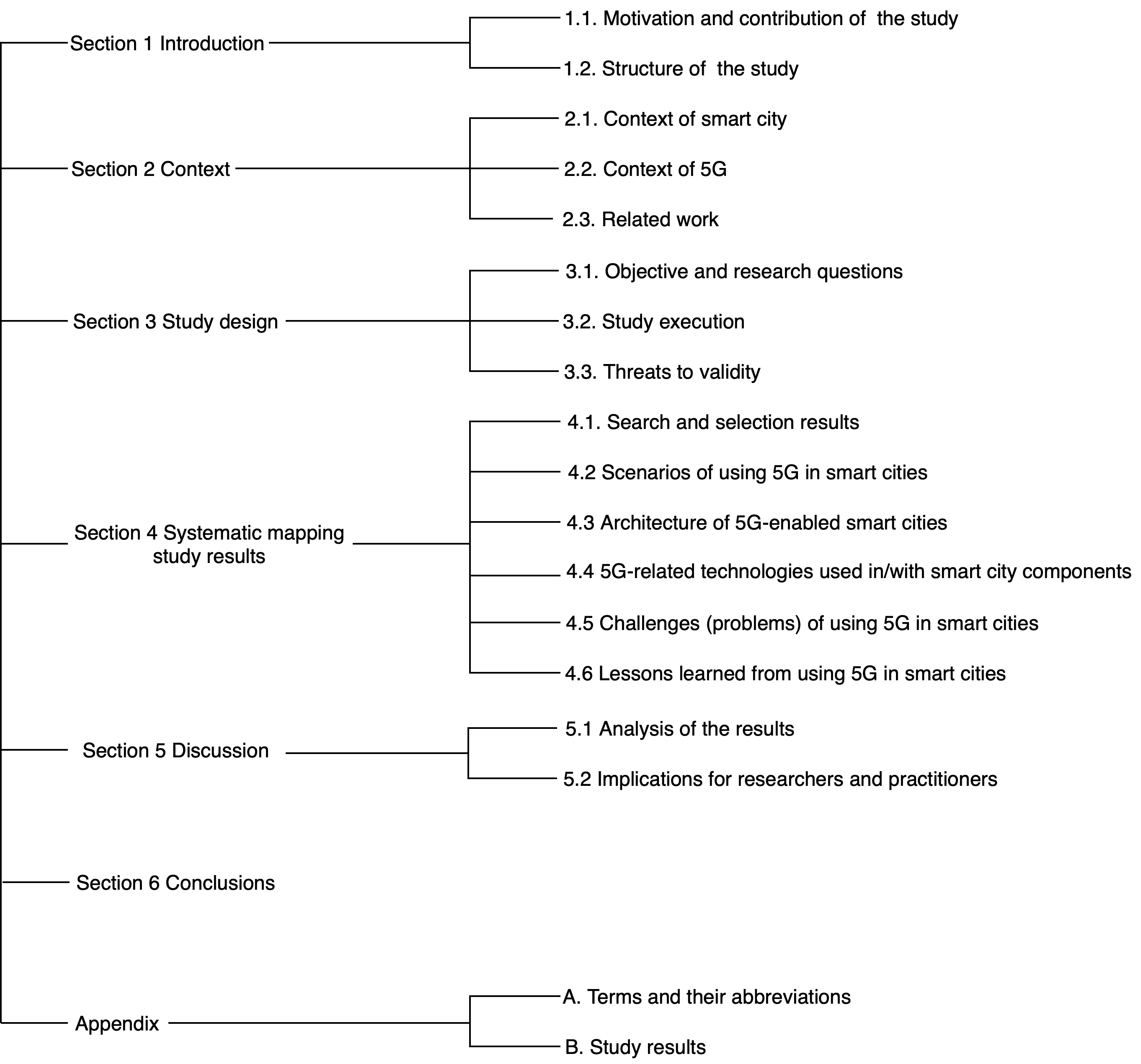}
\caption{Taxonomy of the paper}
\label{Taxonomy_of_the_paper}
\end{figure*}

The rest of the paper is organized as follows. Section \ref{section2} introduces the background information of the mapping study. Section \ref{section3} describes the objective, RQs, study execution, and threats to the validity of the SMS. The results of the study search and selection and the answers to the RQs are provided in Section \ref{section4}. Section \ref{section5} presents the discussion of the results. Section \ref{section6} summarizes this SMS. The taxonomy of the paper is provided in Fig. \ref{Taxonomy_of_the_paper}.

\section{Context}\label{section2}
In this section, we first discuss the topic of smart city regarding the concept and characteristics. Then we describe the topic of 5G regarding 5G technologies and applications. Finally, we compare the related work (i.e., literature reviews) with our mapping study.

\subsection{Context of smart city}
Smart city is not a new but fuzzy concept \cite{hollands2008will}. The concept is usually used in different context that is not always consistent \cite{nam2011conceptualizing}. Besides the definition adopted in this study, we also found several other descriptions of a smart city. Examples are listed as follows. Washburn \textit{et al.} \cite{washburn2009helping} emphasized the use of smart computing technologies as the core of a smart city, providing different services such as administration, education, and healthcare. Giffinger \textit{et al.} \cite{giffinger2007smart} considered that a smart city should have self-decisive, independent, and aware citizens, as well as perform well in governance, mobility, economy, people, living, and environment. Hall’s \cite{hall2000vision} concern regarding a smart city is the management (e.g., monitoring) of critical infrastructures such as buildings, roads, and power, in order to maximize services to citizens, improve city security, make better plans, and optimize resource usage. Harrison \textit{et al}. demonstrated that a smart city includes three aspects, i.e., instrumentation, interconnection, and intelligence \cite{harrison2010foundations}. Rios \cite{rios2012creating} considered smart city as a city that provides inspiration, shares culture, knowledge, and life, and motivates the residents in the city to create and develop their own lives. Partridge \cite{partridge2004developing} emphasized the importance of using Information and Communication Technologies (ICTs) to strengthen the freedom of speech and the accessibility to information and services. Though there are different descriptions regarding the smart city concept, the studies are consistent on three major characteristics of a smart city \cite{alawadhi2012building}\cite{carli2013measuring}: (1) Effectiveness (i.e., providing services to the residents and organizations of a city in an effective way), (2) Environmental benefits (i.e., improving the quality of environment in a city, including energy consumption and pollution), and (3) Innovation (i.e., improving the quality of main components in a city through cutting-edge technologies in order to provide better services to the residents and organizations). 

To be more specific, we listed five smart city dimensions as follows. \textbf{Environment} \cite{arroub2016literature}\cite{chamoso2018tendencies} refers to managing environmental infrastructures (including natural resources and energy) to increase the sustainability of a city. Examples are pollution control, management of sewers and green spaces, and management of grids, lighting, and renewable energies. \textbf{Governance} \cite{arroub2016literature}\cite{chamoso2018tendencies} refers to using strategies, especially through using ICTs, to improve city governing activities (e.g., cooperation among stakeholders). Examples are e-governance, e-democracy, and transparency. \textbf{Living} \cite{arroub2016literature}\cite{chamoso2018tendencies} refers to intelligent ways of living to improve productivity and comfort of residents in a city through technologies (e.g., IoT). Examples are entertainment, public safety, housing quality, and healthcare. \textbf{Mobility} \cite{arroub2016literature}\cite{chamoso2018tendencies} refers to managing traffic problems (e.g., transport congestion) in the context of increasing of population and mobility complexity in cities. Examples are city logistics, mobility information, and mobility of people. \textbf{Economy and society} \cite{arroub2016literature}\cite{chamoso2018tendencies} refers to innovative, digital, competitive, green, socially responsible management regarding economy and society. Examples are innovation and entrepreneurship, culture heritage management, and digital education.

\subsection{Context of 5G}
\begin{figure*}[!t]
\centering
\includegraphics[width=5in]{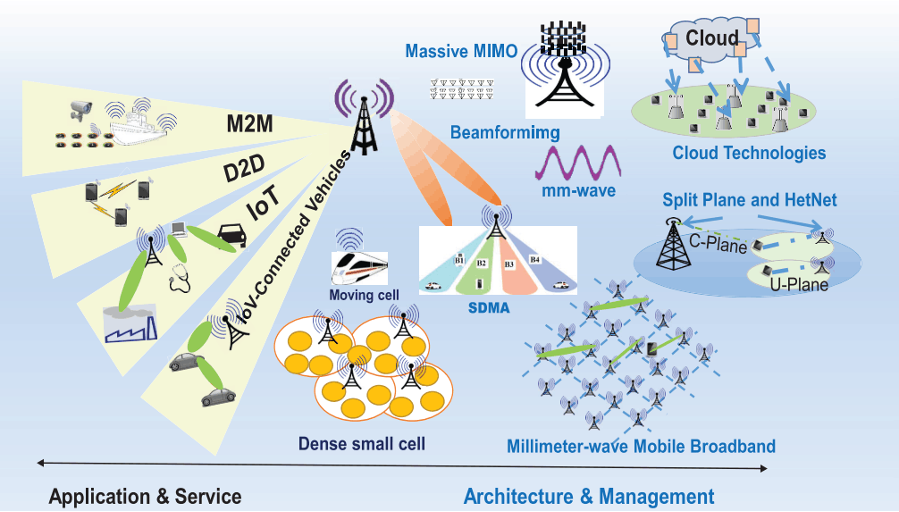}
\caption{A schematic diagram of 5G \cite{agiwal2016next}. 5G is expected to fulfill a set of requirements (e.g., data rates and bandwidth) and be used to support different applications (e.g., Device to Device communications and Machine to Machine communications). It includes many architectural design decisions (e.g., density of small cells) and technologies (e.g., mm-wave and massive MIMO).}
\label{Schematic_diagram_of_5G}
\end{figure*}

5G is expected to satisfy various requirements regarding latency, data rates, bandwidth, coverage, connectivity, and energy consumption \cite{agiwal2016next}\cite{gupta2015survey}. According to literature (e.g., \cite{andrews2014will}\cite{intelligence2014understanding}\cite{chen2014requirements}\cite{minoli2019practical}), the major requirements of 5G include: high data rates (e.g., 1-10 Gbps), low latency (e.g., 1 ms), high bandwidth, massive connected devices (e.g., 100 devices/$m^2$), high availability (e.g., 99.99\%), wide and reliable coverage for anytime and anywhere connectivity, reduction in energy usage (e.g., 90\%), high battery life, and pragmatic deployment cost and service price. There are a set of key architectural design decisions and technologies of 5G \cite{agiwal2016next}\cite{gupta2015survey}\cite{ge2016ultradense}, including mm-wave, massive Multiple Input Multiple Output (MIMO), beamforming, full duplex, Polar/Low-density Parity-check (LDPC) coding, Orthogonal Frequency Division Multiplexing (OFDM), Software Defined Network (SDN), Network Functions Virtualization (NFV), network slice, edge computing, the Cloud Radio Access Network (C-RAN, sometimes referred to as Centralized Radio Access Network) architecture, the Distributed Radio Access Network (D-RAN) architecture, different types of cells (i.e., macrocells, microcells, picocells, and femtocells), and ultra-dense network. Agiwal \textit{et al.} \cite{agiwal2016next} described 5G using a schematic diagram as shown in Fig. \ref{Schematic_diagram_of_5G}. Below we describe three examples. 

(1) The performance of wireless communication is related to many factors, such as spectral efficiency, bandwidth, and cells \cite{stallings2007data}. Current wireless communication usually uses the spectrum between 300 MHz and 3 GHz band, which is identified as the sweet spot \cite{andrews2014will}\cite{rappaport2014wireless}. Though the spectrum is rather reliable for propagation over several kilometers in different context, it may not be able to address the exploding communication traffic (e.g., from mobile phones) and connectivity problems (e.g., large amount of devices connected in a dense area) \cite{khan2012millimeter}. As the first step, instead of using mm-wave (which will eventually be utilized later on), many countries developed sub-6 GHz band (i.e., still using centimeter wave) for 5G communication. The reason is because of the unsolved problems, i.e., higher frequencies are more subject to weather impairments, cannot travel longer distances and is more difficult to penetrate building walls compare to lower frequencies \cite{minoli2019practical}. If we can solve the problems, 5G communication can use mm-wave band up to 300 GHz. Note that only 57 GHz - 64 GHz band and 164 GHz - 200 GHz band are unsuitable for communications \cite{agiwal2016next}. 

(2) In the 5G context, different types of 5G cells (i.e., macrocells, microcells, picocells, and femtocells) coexist \cite{al20195g}. Microcells, picocells, and femtocells are also called small cells. Because of the densely deployment nature, 5G network contains many servers, routers, and other types of devices. Configure and maintain these devices is a challenge in 5G network. SDN offers the swiftness and flexibility to address the configuration and maintenance challenge through splitting the management of 5G network into control plane and data plane \cite{lai2015buffer}\cite{agyapong2014design}. Besides decoupling the network control, SDN enables the network control to be programmable and makes the infrastructure (e.g., hardware) be abstract for applications and services \cite{nizzi2018evaluation}. Interaction between the control plane and data plane can be implemented through open interfaces (e.g., Open Flow) \cite{arslan2015software}. 

(3) Another example is the C-RAN architecture, which aims to address problems regarding, for example, high data rates \cite{checko2014cloud}. In C-RAN, there are three components, i.e., baseband pool, front haul, and remote units (i.e., Remote Radio Head, RRH / Remote Radio Unit, RRU or Active Antenna Unit, AAU) \cite{agiwal2016next}. Baseband resources are virtualized and pooled at Baseband Unit (BBU), which are not deployed with cell sites, but at one or more remote places \cite{cvijetic2014optical}. Remote units are connected to baseband pool through fiber \cite{cvijetic2014optical}. Though the overall 5G network architecture is complicated (e.g., different types of cells, MIMO, and complex structure of heterogeneous networks), based on virtualization, SDN, and other technologies, C-RAN is able to improve the overall quality of the 5G network architecture, mobility, energy consumption, coverage performance, among others. \cite{checko2014cloud}. Note that to fulfill different requirements in 5G communication, there are several definitions and classifications of the components, their connections, and deployments in a 5G C-RAN architecture (i.e., different architectural design decisions). As an example, instead of using BBUs, we can also divide a BBU into a Centralized Unit (CU) and a Distributed Unit (DU). Moreover, regarding the deployment, CU can be deployed with DU (i.e., CU and DU are at the same place), but can also be deployed separately.

Applications of 5G include many aspects, such as Device to Device (D2D) communications, Machine to Machine (M2M) communications, IoT, Vehicle to Everything (V2X), transportation (e.g., high speed trains), healthcare, financial (e.g., payments), smart city, and smart grid \cite{agiwal2016next}. These scenarios have various components, leading to different system architecture, technologies, challenges, and lessons learned of using 5G. 

\subsection{Related work}
There are many literature reviews related to our topic. Mustakim focused on 5G network architecture and emerging technologies that enable vehicular network for smart vehicles in smart cities \cite{ha2020vehicular}. Mustakim briefly introduced Vehicle to Vehicle Communication (V2V), Vehicle to Pedestrian Communication (V2P), Vehicle to Infrastructure Communication (V2I), and intra vehicle communication. Then they analyzed vehicle networking communication standards and 5G vehicular networks. Qadir \textit{et al.} reviewed papers regarding path planning for unmanned aerial vehicles (UAVs) and 5G communications in the field of disaster management in smart cities \cite{qa2021addressing}. They initially identified 139 studies from the last decade and finally included 36 studies in their review. Ullah \textit{et al.} analyzed studies related to applications of AI and machine learning in smart cities, including a topic of deep reinforcement learning based UAV applications in 5G communication \cite{ul2020applications}. They emphasized more on the UAVs role in 5G communication which can further play a critical role in smart city creation and sustainability, including deep reinforcement learning based UAVs-assisted mm-wave communication and UAV positioning for throughput maximization and data offloading. Wang \textit{et al.} reviewed the field of non-terrestrial wireless technologies for smart city IoTs. In their study, they also summarized the case of using UAV with 5G and IoT in smart cities \cite{wang2020nonterrestrial}. Mistry \textit{et al.} presented an in-depth survey of state-of-the-art proposals having 5G-enabled IoT as a backbone for blockchain-based industrial automation for the applications, such as smart city, industry, supply chain, and agriculture \cite{mi2020blockchain}. In the field of smart city, they analyzed and compared seven studies published between 2014 and 2018. Haris and Al-Maadeed also focused on using blockchain technology in 5G enabled IoT, including security, performance, network issues in the IoT devices based on 5G technology and integration of blockchain technology in 5G-IoT devices \cite{ha2020integrating}. In their work, smart city was considered as a blockchain application in 5G enabled IoT. Kasznar \textit{et al.} conducted a review on multiple dimensions of smart cities' infrastructure \cite{ka2021multiple}. They aimed to answer three questions: (1) What are the various dimensions on which smart city infrastructure research focuses? (2) What are the themes and classes associated with these dimensions? (3) What are the main shortcomings in current approaches, and what would be a good research agenda for the future? They selected 75 studies out of more than 12,000 studies. In their review, they used three dimensions of the smart city concept, i.e., technology, institution, and community, and 5G is treated as a subject in the technology dimension. Zhao \textit{et al.} performed a review on nanogenerators for smart cities using 5G and IoT \cite{zhao2021nanoge}. Specifically, their study shows how nanogenerators can be a game changer in the development of smart cities under 5G services, including smart healthcare, human machine interface, smart vehicles, intelligent transportation, wind-based energy, and water-based energy. Huseien and Shah conducted a scoping review to analyze the benefits of 5G to enhance the efficiency of smart cities and minimize climate change impacts in Singapore \cite{hu2021potential}. Their results reveal that the smart management of energy, wastes, water resources, agriculture, risk factors, and the economy adopted in Singapore can remarkably contribute to reducing climate change, thus attaining the sustainability goals. Miladić-Tešić \textit{et al.} focused on providing an overview of recent developments of advanced optical networking to provide 5G transport networks and their applications in connecting a huge number of devices in future smart city infrastructures \cite{mi2021optical}. Their study covers four RQs: (1) What are the requirements of 5G wireless on optical networking? (2) What are the available and emerging optical technologies supporting 5G? (3) What are the benefits of flexibility in optical domain supporting 5G communication infrastructure? (4) What are the research challenges on network flexibility implementation?

Though there are many literature reviews related to our topic, we see the following limitations: (1) the focus of some related work is not using 5G in smart cities, but some other related fields. Examples are using blockchain in smart cities (5G is only a small part of the whole system), IoT in smart cities (5G is only a small part of the whole system), and applications of 5G (smart city is only one example of using 5G). This usually leads to the result that the related work talks 5G in smart cities without details. (2) the research scope is limited. For example, some studies only focus on a specific part of smart cities. (3) Certain related work does not follow a systematic process to search, select, and analyze studies, and therefore, it may lead to threats, such as omitting primary studies and decreasing the quality of the extracted data from the studies because of personal bias. To fill the gap of the exiting work, we decided to conduct an SMS that covers the literature published between January 2012 and December 2019 regarding using 5G in smart cities. The scenarios, architecture, technologies, challenges, and lessons learned of using 5G in smart cities are summarized and further analyzed based on the selected studies.

\section{Study design}\label{section3}
There are many types of empirical methods employed in empirical studies \cite{shull2008guide}\cite{wohlin2012experimentation}, including experiment, survey, case study, SLR, and SMS. In the field of empirical study research, an experiment is used to examine cause-effect relationships between different variables characterizing a phenomenon \cite{wohlin2012experimentation}. A survey is a system for collecting information from or about subjects (e.g., people and projects) to describe, compare, or explain their knowledge, attitudes, and behavior \cite{fink2003thesurvey}. Sometimes researchers also use the term “survey” for a study that is actually a type of literature review. A case study is an empirical inquiry that draws on multiple sources of evidence to investigate one or a small number of instances of a contemporary phenomenon within its real-life context, especially when the boundary between phenomenon and context cannot be clearly specified \cite{runeson2012casestudy}.
SLRs, literature surveys, and SMSs are secondary studies. A secondary study does not generate any data from direct observation or measurement, instead, it analyses a set of primary studies and usually seeks to aggregate the results from these studies to provide stronger forms of evidence about a particular phenomenon \cite{kitchenham2015evidence}. An SLR usually provides an objective and unbiased approach to find relevant primary studies for extracting, aggregating, and synthesizing the data from these studies \cite{kitchenham2015evidence}.
The objective of literature surveys is similar to SLRs. However, literature surveys usually do not follow a systematic process to search, select, and analyze included papers.
An SMS is a type of secondary study that can be used to systematically extract and compile knowledge from primary studies \cite{petersen2015sms}. Both literature surveys, SLRs, and SMSs are commonly used in many research fields (e.g., medical research \cite{ouhbi2020connected}\cite{sadoughi2020internet}\cite{siddiqi2006occurrence} and computer science research \cite{agiwal2016next}\cite{dybaa2008empirical}\cite{petersen2015guidelines}).

As explained in Section \ref{section1}, we decided to conduct an SMS on the topic of using 5G in smart cities. In this section, we introduce the objective, RQs, study execution, and threats to the validity of the SMS.

\subsection{Objective and research questions}
The objective of the SMS is to explore and analyze the state of the art of using 5G in smart cities. We further decomposed the objective into five RQs (see Table \ref{Research questions and their rationale}). 

\begin{table*}[h]
\centering
\caption{RQs and their rationale. The RQs can be mapped to the contribution presented in Section 1.}
\label{Research questions and their rationale}
\begin{tabular}{|p{0.3\columnwidth}|p{0.6\columnwidth}|}
\hline
\textbf{Research question} &
  \textbf{Rationale} \\ \hline
{RQ1: What are the scenarios of using 5G in smart cities?} &
  {There are many working areas related to the topic of the SMS. The answer of this RQ helps researchers and practitioners to understand the current focuses of using 5G in smart cities.} \\ \hline
{RQ2: What is the architecture of 5G-enabled smart cities?} &
  {Through analyzing the relationships between 5G and specific smart city components, and the mapping of 5G-related technologies to those components, this RQ intends to understand the architecture of using 5G in smart cities via a systematic viewpoint.} \\ \hline
{RQ3: Which 5G-related technologies can be used in/with smart city components?} &
  {5G is composed of a set of technologies, such as mm-wave, beamforming, Massive MIMO, radio access, SDN, and NFV \cite{agiwal2016next}. The answer of this RQ identifies the major 5G-related technologies used in the smart city context.} \\ \hline
{RQ4: What are the challenges (problems) of using 5G in smart cities?} &
  {Through identifying the challenges (problems) of using 5G in smart cities, this RQ helps researchers to form new research directions in this area and practitioners to be aware of the weak points of using 5G in smart cities.} \\ \hline
{RQ5: What are the lessons learned from using 5G in smart cities?} &
  {Based on the previous experience of using 5G in smart cities, this RQ helps researchers and practitioners to come up with better solutions in this area.} \\ \hline
\end{tabular}
\end{table*}

\subsection{Study execution}
This section introduces the study execution process, study search and selection, and data extraction and analysis.

\subsubsection{Study execution process}
\begin{figure*}[!t]
\centering
\includegraphics[width=5in]{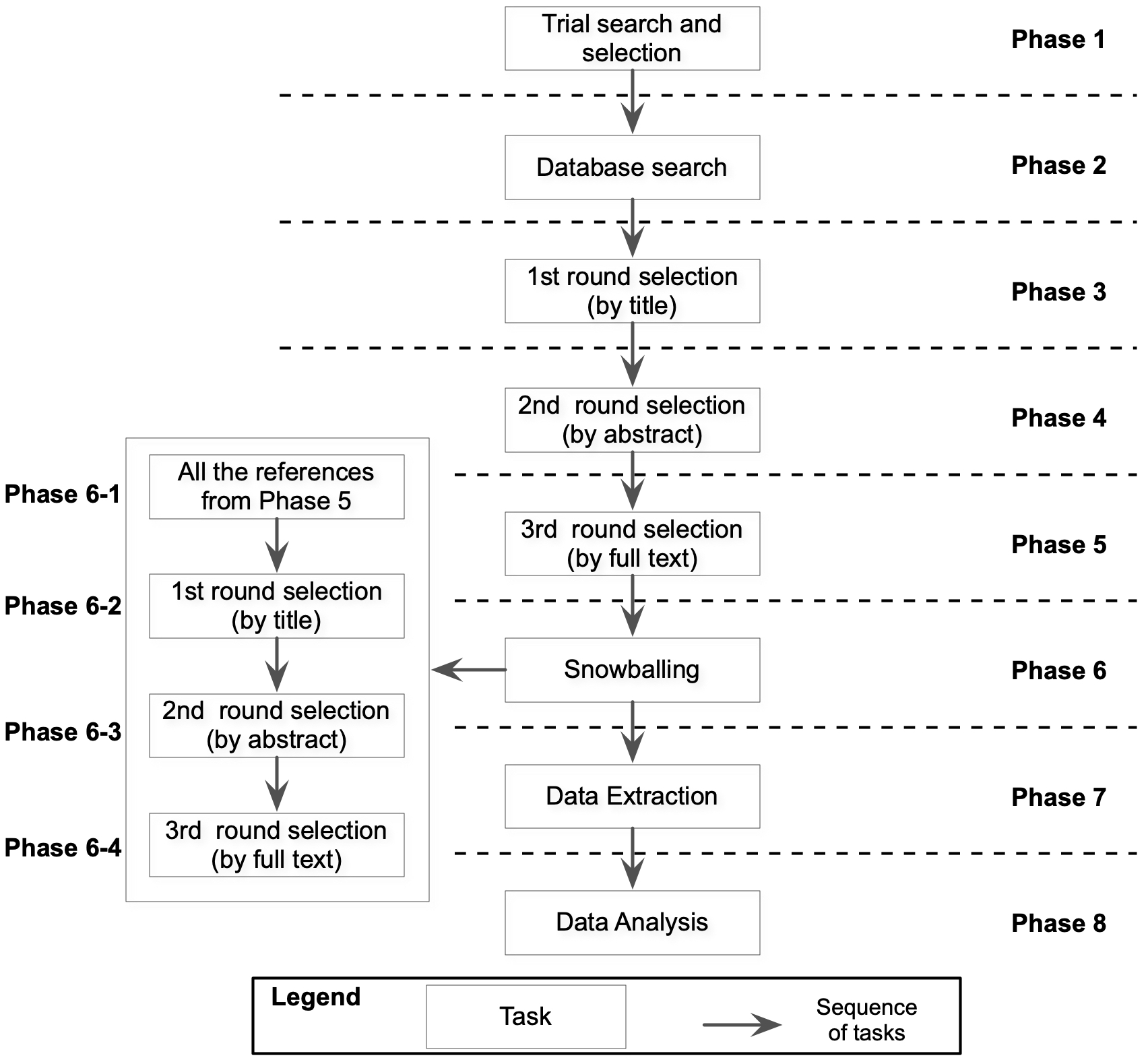}
\caption{Study execution process. We followed a systematic process which is composed of eight phases to search, select, and analyze relevant studies on using 5G in smart cities.}
\label{Study_execution_process}
\end{figure*}

The study execution process is shown in Fig. \ref{Study_execution_process}, which is composed of eight phases. We identified the preliminary search terms from related literature. Since different databases offered various search engines and capabilities, we used the identified search terms and conducted a trial search and selection. Results and problems were discussed by the authors. We refined the search strategy learned from the trial search, and a formal search was conducted to search potentially relevant primary studies in the databases. Then we performed three rounds of study selection through title, abstract, and full text, respectively. All the results were recorded during the search and selection process. After Phase 5, we conducted a manual search (i.e., Phase 6-1) through using the snowballing technique \cite{wohlin2014guidelines} based on the outputs from Phase 5. Then we used a 3-step process (i.e., Phase 6-2, 6-3, and 6-4) for further selection. Finally, data was extracted and analyzed in Phase 7 and 8, respectively.
We started writing the protocol of this work in December 2019, and finished the design, data collection, and data analysis in July 2020.

\subsubsection{Study search and selection}
Based on our knowledge and related literature (e.g., \cite{dybaa2008empirical}\cite{ali2010systematic}), we chose seven databases as the sources for database search in the SMS. The selected databases are representative as they are used as the sources for study search in many related SMSs and SLRs. Since the search engines of the databases were different, we listed the name and search area of each database as follows: (1) ACM Digital Library (title and abstract), (2) IEEE Explore (metadata that includes title, abstract, and indexing terms), (3) Science Direct (title, keywords, and abstract), (4) SpringerLink (full text), (5) Wiley InterScience (title and abstract), (6) EI Compendex (subject, title, and abstract), and (7) ISI Web of Science (topic that includes title, abstract, author keywords, and keywords plus). Note that the ISI Web of Science database was renamed as Clarivate Analiytics in 2018. We also conducted an additional search on the Scopus database with the same settings as used for the seven databases. However, all the papers retrieved from the Scopus database were included in the search results of the seven databases. Therefore, we did not include the Scopus database in this study. Moreover, we found that industry and policy started analyzing and discussing requirements of 5G in 2012 \cite{harborth2017standardization}. Therefore, we set the time period between January 2012 and December 2019 in database search. 
The trial search was conducted in January 2020, and the formal search was conducted in February 2020.

Nam and Pardo \cite{nam2011conceptualizing} conducted a study on conceptualizing smart city, and came up with a set of definitions and conceptual relatives. Therefore, the search terms regarding smart city used in this study are based on their research \cite{nam2011conceptualizing}. We used Boolean ``OR'' to join alternate words and synonyms and Boolean ``AND'' to join major terms. The following search query was used in database search:

\textit{(5G) AND (``smart city'' OR ``smart cities'' OR ``digital city'' OR ``digital cities'' OR ``intelligent city'' OR ``intelligent cities'' OR ``ubiquitous city'' OR ``ubiquitous cities'' OR ``wired city'' OR ``wired cities'' OR ``hybrid city'' OR ``hybrid cities'' OR ``information city'' OR ``information cities'' OR ``creative city'' OR ``creative cities'' OR ``learning city'' OR ``learning cities'' OR ``humane city'' OR ``humane cities'' OR ``knowledge city'' OR ``knowledge cities'' OR ``smart community'' OR ``smart communities'')}

To select studies from the retrieved results from database search, we used the following inclusion criteria:

I1: The paper concerns using 5G in smart cities.

We used the following exclusion criteria:

E1: The paper concerns 5G or smart city separately. 

E2: The paper talks using 5G in smart cities without details, i.e., no data can be extracted to answer the RQs.

E3: The paper is gray literature (i.e., not peer-reviewed, such as technical report and work in progress) \cite{keele2007guidelines}.

E4: The paper is not written in English.

E5: If two papers publish the same study in different venues (e.g., conference and journal), the less mature one is excluded.

E6: The paper is a secondary study (e.g., literature reviews and surveys).

E7: The paper is not available in full-text.

We also found certain topics that are usually related to the smart city topic, such as automated driving, smart grid, and smart factory. However, unless the context of the study is smart city (e.g., automated driving in a smart city), the study was not considered (i.e., not qualified for I1). 

\subsubsection{Data extraction and analysis}
Table \ref{Data items, their description, and related RQs} shows the data items, their description, and related RQs extracted after study search and selection. The information was recorded through the MAXQDA tool. The length of the extracted data can be various (e.g., one sentence, several sentences, one paragraph, or several paragraphs). We used both quantitative and qualitative analysis (i.e., descriptive statistic and Constant Comparison) based on the RQs. Constant Comparison provides a systematic way to generate incidents, concepts, and categories from the extracted data, and a continuous process of verification of the generated incidents, concepts and categories \cite{glaser1968discovery}.

\begin{table*}[h]
\centering
\caption{Data items, their description, and related RQs. ``\textit{ID}'', ``\textit{Title}'', ``\textit{Author list}'', ``\textit{Year}'', ``\textit{Venue}'', and ``\textit{Evidence level}'' are common information when conducting SMSs. Other data items can be mapped to the RQs.}
\label{Data items, their description, and related RQs}
\begin{tabular}{|p{0.2\columnwidth}|p{0.6\columnwidth}|p{0.1\columnwidth}|}
\hline
\textbf{Data item} & \textbf{Description} & \textbf{RQ} \\ 
\hline
ID & The identity of the study & N/A \\ 
\hline
Title & The title of the study & N/A         \\ 
\hline
Author list & The full name of the authors & N/A         \\ 
\hline
Year & The year when the study was published & N/A         \\ 
\hline
Venue & {The name and type (i.e., journal, conference, and book) of the venue where the study was published} & N/A \\ 
\hline
Evidence level & {The evidence is regarding how the proposed 5G architecture or technologies were evaluated in the study. We follow the evidence hierarchy in six levels proposed in \cite{alves2010requirements}.} & N/A \\ 
\hline
Scenario & The scenarios of using 5G in smart cities & RQ1         \\ 
\hline
Architecture & The architecture of 5G-enabled smart cities & RQ2         \\ 
\hline
5G technology & {The 5G-related technologies that can be used in/with smart city components, and how they are used in/with smart city components} & RQ3 \\ 
\hline
Challenge & The challenges (problems) of using 5G and 5G-related technologies in smart cities & RQ4         \\ 
\hline
Lesson learned & The lessons learned from using 5G in smart cities & RQ5         \\ 
\hline
\end{tabular}
\end{table*}

\subsection{Threats to validity}
We discussed the threats to the validity of the SMS according to study search and selection and data extraction and analysis.
One threat is that we may have omitted studies in database search because of improper search sources, search terms, and search strategies. To reduce the threat, we chose 7 popular databases and search terms according to our knowledge and related literature. We performed snowballing to manually search and select studies. We also conducted a trial search to develop the most suitable search strategy (e.g., search area of the database).

Due to the subjective nature of study selection, another threat is that we may have incorrectly included or excluded studies in this SMS. To reduce the threat, besides conducting snowballing and a trial selection in the databases, we also carefully designed the selection criteria. Moreover, all the selection phases were performed by two authors independently and the other authors reviewed the results. Conflicts were addressed by all the authors through meetings.

Personal bias and mistakes may have a negative impact on data extraction and analysis. To address the problem, the data extraction and analysis were done by one author, and the results were reviewed by another author. Conflicts were addressed by all the authors through meetings. We also used the MAXQDA tool in data extraction and analysis. The tool offered functions such as coding and visualization, which helped us to partially reduce mistakes.

\section{Systematic mapping study results}\label{section4}

In this section, we provide the results of the study search and selection and the answers to the RQs. Note that ``\textit{the studies}'' in this section means the selected 32 studies of the SMS.
\subsection{Search and selection results}
This section provides the search and selection results. As shown in Fig. \ref{Search_and_selection_results_of_the_studies} in the appendix, we collected 1366 studies from the seven databases. For the three-round selection process, we selected 206 studies, 83 studies, and 31 studies, respectively. 
Moreover, one study was identified through snowballing, and consequently 32 studies were finally included in this work, as shown in Table \ref{tab:long}.

\begin{center}
\begin{longtable}{|p{0.84\columnwidth}|p{0.06\columnwidth}|}
\caption{Title and publication year of the selected studies.} \label{tab:long} \\

\hline \multicolumn{1}{|c|}{\textbf{Title}} & \multicolumn{1}{c|}{\textbf{Year}} \\ \hline 
\endfirsthead

\hline \multicolumn{1}{|c|}{\textbf{Title}} & \multicolumn{1}{c|}{\textbf{Year}} \\ \hline 
\endhead

\hline
\endlastfoot
{UAV Network and IoT in the Sky for Future Smart Cities} & 2019 \\ \hline
{The technical concept of using the 700 MHz band as a base for 5G smart cities networks in Poland} & 2018 \\ \hline
{Softwarization and virtualization in 5G networks for smart cities} & 2016 \\ \hline
{Smart mobile city services in the 5G era} & 2018 \\ \hline
{Smart home and smart city solutions enabled by 5G, IoT, AAI and CoT services} & 2014 \\ \hline
{Smart city traffic monitoring system based on 5G cellular network, RFID and machine learning} & 2019 \\ \hline
{Smart city as a 5G ready application} & 2018 \\ \hline
{Scenario planning for 5G light poles in smart cities} & 2017 \\ \hline
{A real case of implementation of the future 5G city} & 2019 \\ \hline
{The INCIPICT project and the 5G trial: A living lab for the city of L'Aquila} & 2018 \\ \hline
{Impact of 5G technologies on smart city implementation} & 2018 \\ \hline
{Evaluation of IoT and videosurveillance applications in a 5G smart city: The Italian 5G experimentation in Prato} & 2018 \\ \hline
{Enabling technologies to realise smart mall concept in 5G era} & 2018 \\ \hline
{Efficient media streaming with collaborative terminals for the smart city environment} & 2017 \\ \hline
{Dynamic traffic prediction with adaptive sampling for 5G HetNet IoT applications} & 2019 \\ \hline
{Distributed network infrastructure for community services in the framework of next generation mobile networks} & 2017 \\ \hline
{Development and implementation of a smart city use case in a 5G mobile network's operator} & 2017 \\ \hline
{Deploying 5G-technologies in smart city and smart home wireless sensor networks with interferences} & 2015 \\ \hline
{Data traffic model in machine to machine communications over 5G network slicing} & 2016 \\ \hline
{Cost model for a 5G smart light pole network} & 2019 \\ \hline
{A context-aware radio access technology selection mechanism in 5G mobile network for smart city applications} & 2019 \\ \hline
{Compute and network virtualization at the edge for 5G smart cities neutral host infrastructures} & 2019 \\ \hline
{Compact base station antenna based on image theory for UWB/5G RTLS embraced smart parking of driverless cars} & 2019 \\ \hline
{City strategies for a 5G small cell network on light poles} & 2019 \\ \hline
{5G Wireless Micro Operators for Integrated Casinos and Entertainment in Smart Cities} & 2018 \\ \hline
{5G smart city vertical slice} & 2019 \\ \hline
{5G optimized caching and downlink resource sharing for smart cities} & 2018 \\ \hline
{A 5G new smart city services facilitator model} & 2019 \\ \hline
{5G field trials in the smart city and medical service areas toward social implementation of 5G} & 2018 \\ \hline
{5G converged cell-less communications in smart cities} & 2017 \\ \hline
{A business case for 5G mobile broadband in a dense urban area} & 2019 \\ \hline
{5G network architecture, functional model and business role for 5G smart city use case: Mobile operator perspective} & 2018 \\ \hline
\end{longtable}
\end{center}

\begin{figure*}[!t]
\centering
\includegraphics[width=5in]{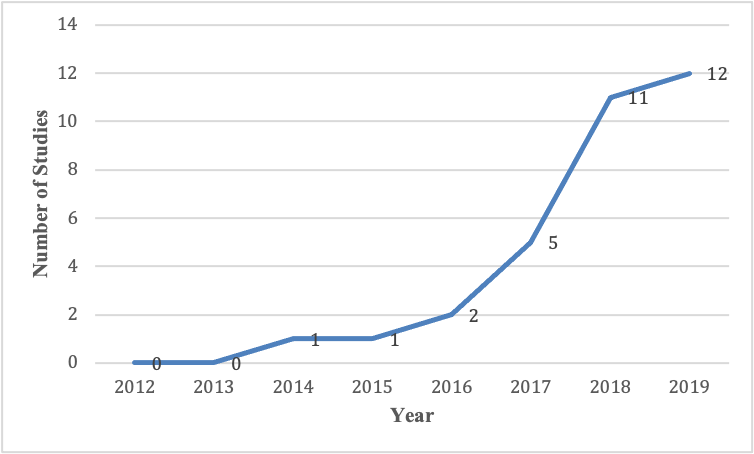}
\caption{Number of studies on using 5G in smart cities over time period (2012-2019). After the searching and selection phases, we identified and collected 32 relevant studies on the topic of using 5G in smart cities. The number increases steadily since 2015, and reaches a peak in 2019.}
\label{Number_of_studies}
\end{figure*}

The sources of the studies on using 5G in smart cities are shown in Table \ref{Sources of the studies on using 5G in smart cities} in the appendix. The 32 studies are published in 27 publication venues, containing 17 conferences (out of 27, 63.0\%), 9 journals (out of 27, 33.3\%), and 1 book (out of 27, 3.7\%). The top 5 venues are IEEE Communications Magazine, Wireless Personal Communications, IEEE Access, Smart Cities \& Information and Communication Technology (CTTE-FITCE), and International Conference on Communications (COMM). 
The number of the studies on using 5G in smart cities over time period is from 0 to 12 as shown in Fig. \ref{Number_of_studies}. The number increases steadily since 2015, and reaches a peak in 2019.

Table \ref{Evidence regarding using 5G in smart cities of the studies} shows the evidence of using 5G in smart cities from the studies. 17 studies (out of 32, 53.1\%) report results based on academic studies and 13 studies (out of 32, 40.6\%) use demonstration or toy examples. Only 2 studies (out of 32, 6.3\%) report using 5G in smart cities based on industrial studies. If a study provides evidence at multiple levels, its highest level is counted. There are studies claiming that their proposed methods had been (or would be) used in a real city project (e.g., the MATILDA project and the H2020 5GCity project). However, we did not find any report from the studies on evaluation of the proposed methods used in the mentioned city project (i.e., no industrial studies or industrial practice reported). Therefore, we did not consider such statements as evidence.

\begin{table*}[htbp]
\centering
\caption{Evidence regarding using 5G in smart cities in the studies. It is usually unclear whether the proposed methods or the obtained results from literature can be used in practice. We used a rigorous assessment method to analyze the evidence level of adopting the proposed methods or the obtained results from the selected studies in practice.}
\label{Evidence regarding using 5G in smart cities of the studies}
\begin{tabular}{|p{0.15\columnwidth}|p{0.6\columnwidth}|p{0.15\columnwidth}|}
\hline
\textbf{Evidence level} & \textbf{Description} & \textbf{Number of studies} \\ 
\hline
Level 1 & {There is no evidence regarding using 5G in smart cities.} & 0 \\ 
\hline
Level 2 & {There is evidence regarding using 5G in smart cities based on demonstration or working out toy examples.} & 13 \\ 
\hline
Level 3 & {There is evidence regarding using 5G in smart cities based on expert opinions or observations.} & 0  \\ 
\hline
Level 4 & {There is evidence regarding using 5G in smart cities based on academic studies.} & 17 \\ 
\hline
Level 5 & {There is evidence regarding using 5G in smart cities based on industrial studies.} & 2 \\ 
\hline
Level 6 & {There is evidence regarding using 5G in smart cities based on industrial practice.} & 0 \\ 
\hline
\end{tabular}
\end{table*}

Moreover, there are many uncertainties mentioned in the studies, such as how 5G will be used in real context and configuration of 5G network in smart cities. One way to deal with the uncertainties is to make assumptions. An assumption is something uncertain but taken for granted (or accepted as true). We found that 16 studies (out of 32, 50.0\%) include assumptions of 5G network design or smart city scenarios. An example is: ``\textit{We assumed that at the beginning of the study period macrocells and small cells were equipped with cross-polarised antennas and the user equipment with 2 antennas.}'' Another example is: ``\textit{The shopping mall test case is formed with four rows of femtocells, assuming each shop there is an access point and a pedestrian corridor in the middle.}''

\subsection{Scenarios of using 5G in smart cities}
This section reports the results of RQ1. 27 studies (out of 32, 84.4\%) demonstrate at least one scenario of using 5G in smart cities. Others (5 studies out of 32, 15.6\%) describe smart city as a whole without presenting any specific scenario. Some mentioned scenarios (e.g., supply chain and logistic, waste recycling, environment sustainability, 3D image, and localization) do not have details in the studies, which are not included in this section. We also found that the line between two scenarios is not always clear. As an example, casinos (belongs to the entertainment scenario) can include certain sort of transportation (e.g., transportation between casinos and train stations). There are no systematic classifications regarding the whole picture of scenarios of using 5G in smart cities from the selected studies. As mentioned in Section 3.2.3, we did not predefine classifications of the scenarios of using 5G in smart cities, but used Constant Comparison to generate incidents, concepts, and categories. This method has also been employed to answer RQ2-RQ5. 

The smart city scenarios are demonstrated in Table \ref{Smart city scenarios}. Transportation (12 studies out of 32, 37.5\%) is the most mentioned scenario in the studies. Note that one study can include zero, one, or multiple scenarios of using 5G in smart cities. We provide three examples as follows. In \cite{liao20185g}, the authors discussed the design and applications of using 5G for casinos and entertainment. The transportation scenario in such context includes various vehicles (e.g., buses, trains, taxis, and shuttles) driving among the casinos, entertainment, gateways (e.g., airports), and other destinations in the city. These vehicles can be autonomous and therefore there is a need to deploy massive sensors all over the city (e.g., vehicles, driveways, and crossroads). 5G can be used for the communications between the sensors, enabling low latency and high reliability to ensure the transportation is safe and foolproof. The 5G network should also be smart, i.e., it should be scalable and flexible to adapt different cases, such as rush crowds getting off work or after a sport game. In \cite{marabissi2019real}, the authors considered that smart transportation should be safe, efficient, reliable, and green. Similar to \cite{liao20185g}, the authors also mentioned that there would be many devices inside or outside vehicles, and 5G can be used for the communications in the V2X network. This includes many use cases, such as road monitoring (i.e., monitoring daily activities and state of the road surface and exchanging data through 5G) and advanced viability of vehicles (i.e., connecting vehicles and control/data centers through 5G to share data). The characteristics of 5G are important in this scenario, i.e., offering real-time communication services as well as supporting high density of network nodes during traffic jam and emergencies. In \cite{sharif2019compact}, the authors described parking in the transportation scenario. They considered that a smart parking system with 5G can (1) help vehicles to find empty parking slots along roadside or underground parking, (2) provide guidance towards the slots, and (3) help to pay for parking fees.

\begin{table*}[htbp]
\centering
\caption{Smart city scenarios of using 5G. The scenarios are scattered across the 32 selected studies, and we used Constant Comparison to analyze and classify these studies. Some mentioned scenarios (e.g., supply chain and logistic, waste recycling, environment sustainability, 3D image, and localization) do not have details in the studies, which are not included. Note that ``\textit{Number of studies}'' means the number of the selected studies (i.e., 32) that mention a specific smart city scenario with 5G.}
\label{Smart city scenarios}
\begin{tabular}{|p{0.2\columnwidth}|p{0.55\columnwidth}|p{0.15\columnwidth}|}
\hline
\textbf{Smart city scenario} & \textbf{Description} & \textbf{Number of studies} \\ 
\hline
Transportation & {Transportation includes management of traffic and accidents on the roads, assisted or autonomous driving, parking regarding both roadside and underground, and related services (e.g., paying fees).} & 12 \\ 
\hline
Public safety & {Public safety refers to safety and security to a city (including the citizens living in the city). Examples are security threat from theft, vandalism, looting, and riots, security for events, and anomalies such as smoke and fire.} & 9 \\ 
\hline
Healthcare & {This is regarding providing healthcare services across metropolitan, regional, and remote locations. Examples are teleradiology, telesurgery, 3D imaging (e.g., brain), emergence care, remote conferencing and patient monitoring, and implanted microchips and sensors.} & 6 \\ 
\hline
City tourism & {City tourism refers to visiting museums, historic buildings, and archaeological sites, as well as others such as malls.} & 5 \\ 
\hline
Entertainment & {In the studies, this refers to entertainment related to casinos (e.g., mega jackpot and anti-counterfeiting lottery).} & 2 \\ 
\hline
Education & {This refers to moving physical class rooms to online and virtue classrooms. Students can learn anytime at anyplace from various online sources. High speed and ultra-low latency are required in this scenario.} & 2 \\ 
\hline
\end{tabular}
\end{table*}

The second smart city scenario is public safety (9 studies out of 32, 28.1\%). The usage of 5G in this scenario is consistent, i.e., supporting data transmissions between cameras and servers. We provide three examples as follows. In \cite{Okumura}, the authors stated that public safety includes services for monitoring and addressing wide area disasters, crowding, dangerous articles, and premeditated crimes. The scenario is composed of many high-resolution cameras scattered in the city or worn by security guards, leading to the need of supporting massive image data transmissions. Enabled by the ultra-high speed and large capacity characteristics, 5G can be used in the scenario. In \cite{nizzi2018evaluation}, the authors also mentioned the necessity of video surveillance for public safety in smart cities. Specifically, they emphasized in some cases, there is a need to get multiple sequences of image frames from different cameras for real-time analysis. Therefore, the required data rate of transmission is much higher than the cases that only need a single image. Using 5G in the public safety scenario can ensure data transmission, and further improve the performance of for example, face recognition, object detection, action detection, and person re-identification. In \cite{rao2018impact}, the public safety scenario includes security threats from theft, riots, and terrorism, leading to the need of real-time video surveillance at for example banks, stores, and roads. The authors pointed out that using 5G in this scenario can fulfill the requirement, enabling face recognition, specific person detection, action detection, among others.

Healthcare (6 studies out of 32, 18.8\%) is in the third position. We provide three examples as follows. In \cite{marabissi2019real}, the healthcare scenario includes the support of a set of health services that can be used remotely, such as telemedicine, telemonitoring, and action detection. The cases of using 5G in the scenario are eMBB, mMTC, and uRLLC. In \cite{dighriri2016data}, the authors focused on constantly monitoring patients with smart M2M devices, for example, gathering information regarding body parameters at home or hospitals. In their smart healthcare system with 5G, they identified five main requirements: (1) sensitive data traffic, (2) 1 ms priority, (3) low latency, (4) 100 Mbps of minimum bandwidth, and (5) support of mobility. In \cite{rao2018impact}, the authors summarized how 5G can support the healthcare scenario: (1) the high bandwidth and massive connected devices characteristics of 5G can support health monitors and consumer devices; (2) the ultra-low latency characteristic of 5G can support remote surgery; (3) the always on connectivity characteristic of 5G can support remote healthcare services being available across metropolitan, regional, and remote locations.

Besides the smart city scenarios, we also identified five smart city infrastructure management scenarios. The core of smart city infrastructure management is to deploy various devices on city infrastructure, and use 5G for the communications between the devices, in order to support different smart city scenarios. For example, in transportation and public safety, there are many use cases regarding 5G-based video cameras. As shown in Table \ref{Smart city infrastructure management scenarios}, 9 studies (out of 32, 28.1\%) focus on lighting management. Video management (also called media management, 8 studies out of 32, 25.0\%) is in the second position, followed by building management (7 studies out of 32, 21.9\%). 

\begin{table*}[htbp]
\centering
\caption{Smart city infrastructure management scenarios. The core of smart city infrastructure management is to deploy various devices on city infrastructure, and use 5G for the communications between the devices, in order to support different smart city scenarios. Note that ``\textit{Number of studies}'' means the number of the selected studies (i.e., 32) that mention a specific smart city infrastructure management scenario with 5G.}
\label{Smart city infrastructure management scenarios}
\begin{tabular}{|p{0.2\columnwidth}|p{0.55\columnwidth}|p{0.15\columnwidth}|}
\hline
\textbf{Smart city infrastructure  management scenario} & \textbf{Description} & \textbf{Number of studies} \\ 
\hline
Lighting management & {Lighting management includes accommodating massive connected actuators and controllers to support not only lighting in streets, but also other city services such as surveillance, advertising, and Wi-Fi hotspots.} & 9 \\ 
\hline
Video management & {This includes management of different types of videos (e.g., surveillance cameras, wearable cameras, cameras equipped on other devices such as drones) for video streaming (e.g., uplink and downlink), information sharing, video traffic management, and real-time video surveillance.} & 8 \\ 
\hline
Building management & {Building management refers to automation in buildings. An example is to constantly monitor the status of a building to detect potential risks and provide specific mechanisms to respond to emergency .} & 7 \\ 
\hline
Energy management & {Energy management supports accurate and real-time analysis of energy usage in different places (e.g., buildings) of a city.} & 4 \\ 
\hline
Network management & {Network management refers to using for example UAVs in the sky to enhance the coverage of the network.} & 2 \\ 
\hline
\end{tabular}
\end{table*}

We further mapped the identified smart city scenarios to the three 5G patterns (i.e., eMBB, mMTC, and uRLLC) as shown in Table \ref{A mapping of the smart city scenarios and smart city infrastructure management scenarios to the 5G scenarios}. For example, for city infrastructure management (e.g., management of buildings and light poles), there would exist a large number of sensors, actuators, and other types of devices that need to be connected in a network (i.e., the mMTC scenario). Note that the network management scenario is mainly regarding the management of components in a 5G network (e.g., network sharing) to improve the performance of the network in smart cities. Therefore, we did not map it to any 5G pattern. 

\begin{table*}[htbp]
\centering
\caption{A mapping of the smart city scenarios and smart city infrastructure management scenarios to the 5G patterns (i.e., eMBB, mMTC, and uRLLC). As an example, regarding using 5G in the Entertainment scenario, the 5G usage patterns include eMBB, mMTC, and uRLLC.}
\label{A mapping of the smart city scenarios and smart city infrastructure management scenarios to the 5G scenarios}
\begin{tabular}{|p{0.6\columnwidth}|p{0.3\columnwidth}|}
\hline
\textbf{Smart city scenario}                           & \textbf{5G patterns} \\ \hline
Entertainment                                          & eMBB, mMTC, uRLLC    \\ \hline
City tourism                                           & eMBB, mMTC           \\ \hline
Education                                              & eMBB, uRLLC          \\ \hline
Healthcare                                             & eMBB, mMTC, uRLLC    \\ \hline
Transportation                                         & eMBB, mMTC, uRLLC    \\ \hline
Public safety                                          & eMBB                 \\ \hline
\textbf{Smart city infrastructure management scenario} & \textbf{5G patterns} \\ \hline
Building management                                    & mMTC                 \\ \hline
Lighting management                                    & mMTC                 \\ \hline
Energy management                                      & mMTC                 \\ \hline
Video management                                       & eMBB                 \\ \hline
Network management                                     & N/A                  \\ \hline
\end{tabular}
\end{table*}

Based on the scenarios, we also identified six types of components of a smart city that are related to 5G. The description of each type is presented in Table \ref{Smart city components related to 5G}. Moreover, specific smart city scenarios (e.g., transportation) may use different components and the connections between the components as well as between 5G and the components could be various.

\begin{table*}[htbp]
\centering
\caption{Smart city components related to 5G. A smart city is composed of a set of components. When using 5G in smart cities, 5G is usually related to different smart city components. This table shows the related smart city components identified from the selected studies.}
\label{Smart city components related to 5G}
\begin{tabular}{|p{0.2\columnwidth}|p{0.7\columnwidth}|}
\hline
\textbf{Component} & \textbf{Description} \\ 
\hline
Technology &{This refers to the technologies used with 5G in smart cities, such as cloud computing, IoT, big data, and AI technologies. These technologies provide different services and cooperate with each other to support a smart city.} \\ 
\hline
Service & {New or adapted services are provided in the context of using 5G in smart cities. Examples are services (e.g., 5G network services and platform services) related to the scenarios identified in Section 4.2.} \\ 
\hline
City infrastructure & {There are various types of city infrastructure related to the usage of 5G according to the scenarios. Examples are fiber, power grid, light pole, building, street, and public space.} \\ 
\hline
Business & {Using 5G in a smart city requires new infrastructure and services, which would bring new business modes, investment and employment opportunities.} \\ 
\hline
Stakeholder & {This includes infrastructure owner, service provider, service operator, and service consumer. Detailed examples are government, developer, network operator, system operator, mobile user, driver, customer, and patient.} \\ 
\hline
Device & {Based on different scenarios and the technologies used, various devices would be employed. Examples are vehicle, server, robot, mobile device, camera, wearable device, and drone.} \\ 
\hline
\end{tabular}
\end{table*}

\subsection{Architecture of 5G-enabled smart cities}
This section reports the results of RQ2. 28 studies (out of 32, 87.5\%) propose and/or discuss the architecture of 5G-enabled smart cities. We identified three types of architecture as follows. Note that one study can describe zero, one, two, or three types of the architecture. An example is \cite{storck20195g}, which describes both a smart city architecture (treating 5G as a component) and a 5G network architecture in smart cities.
\subsubsection{Smart city architecture (treating 5G as a component)}

\begin{figure*}[!h]
\centering
\includegraphics[width=5in]{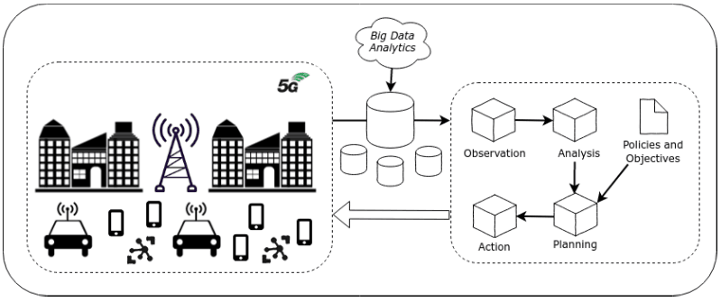}
\caption{Urban computing framework in 5G network \cite{storck20195g}. The 5G part in the architecture is composed of a set of base stations, vehicles, buildings, and mobile devices.}
\label{Urban_computing_framework_in_5G_network}
\end{figure*}

\begin{figure*}[!h]
\centering
\subfigure{}
{
\includegraphics[width=5in]{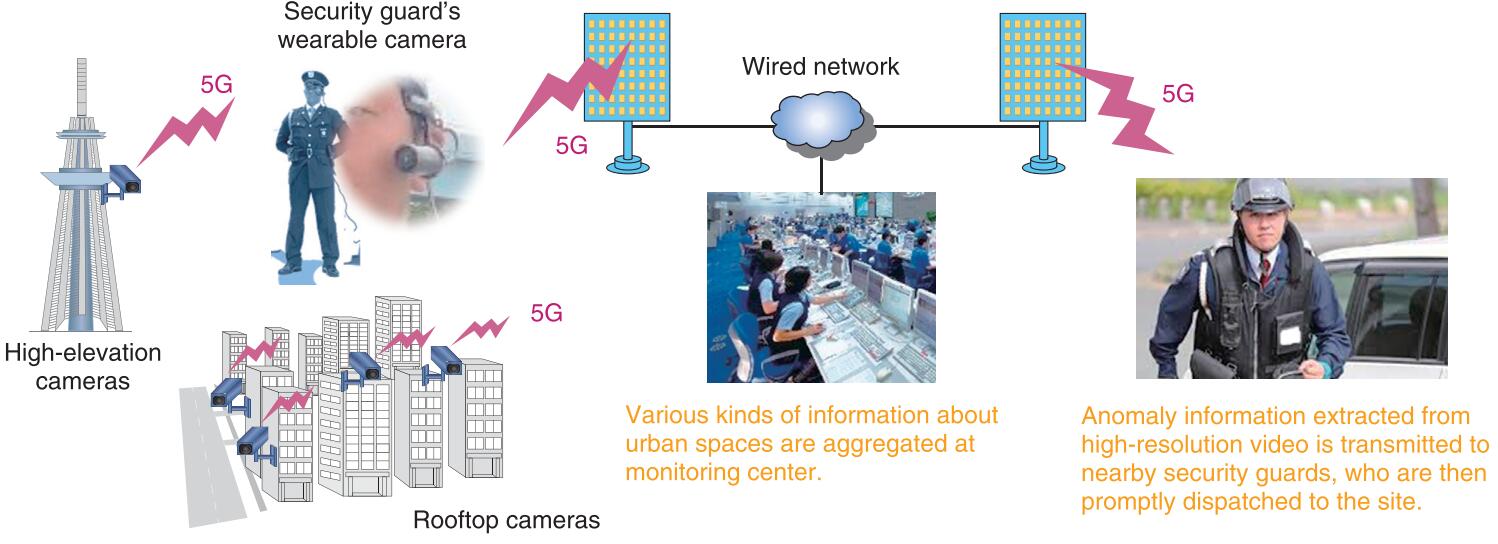}
}

\subfigure{}
{
\includegraphics[width=5in]{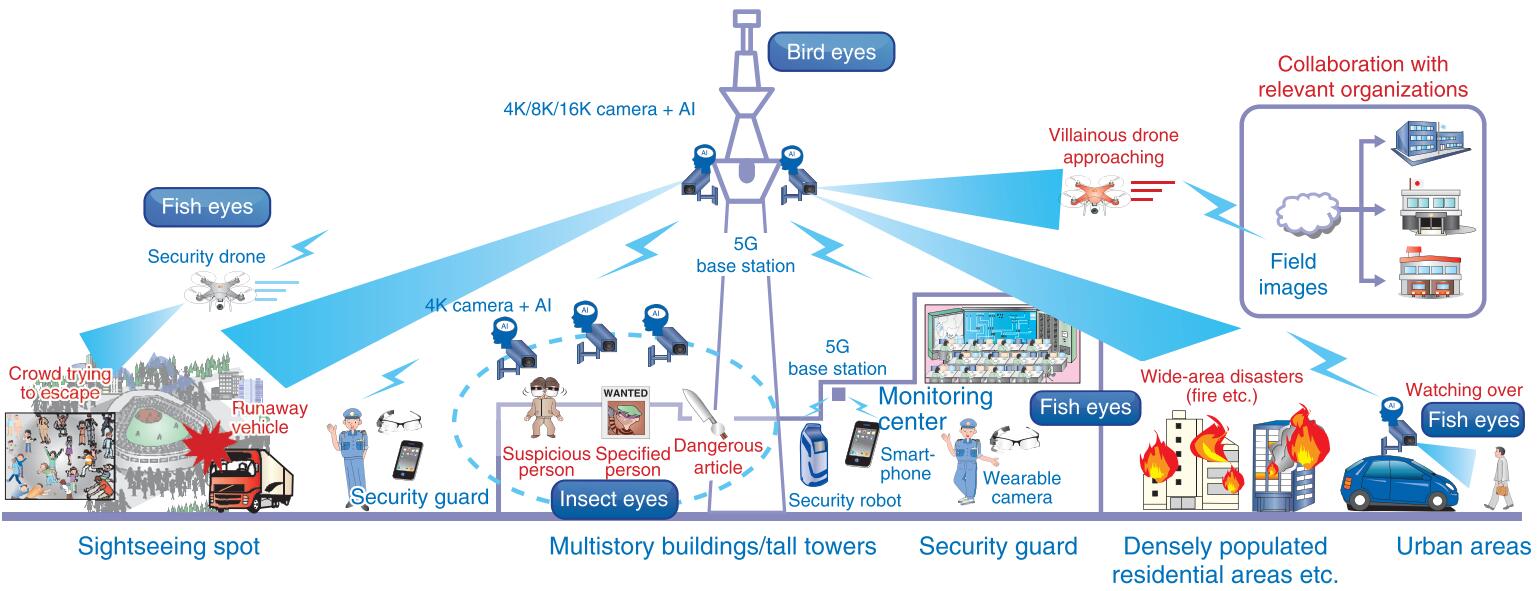}
}
\caption{
\label{smartcityarchitecture_4}
An example of smart city architecture for public safety using 5G \cite{Okumura}. In the architecture, 5G is used for video transmission. The architecture includes various stakeholders, such as security guards, suspicious people, and specified people and devices, such as cameras, robots, smart phones, and drones.
}
\end{figure*}

In this type, 7 studies (out of 32, 21.9\%) provide a general architecture of 5G-enabled smart cities. The architecture treats a smart city as a whole without providing details on specific scenarios or the architecture provided in the studies includes more than one scenario. 
For instance, in \cite{storck20195g}, the authors provided an urban computing framework in 5G network, while the 5G part in the architecture is composed of a set of base stations, vehicles, buildings, and mobile devices (see Fig. \ref{Urban_computing_framework_in_5G_network}). 
For another example, in \cite{dighriri2016data}, the authors provided a 5G network slicing architecture for smart cities that includes multiple applications: usage of mobile devices (e.g., video), smart healthcare, and smart traffic monitoring. 
In \cite{Okumura}, the authors described two examples of smart city architecture using AI and 5G. The architecture includes various stakeholders such as security guards, suspicious people, and specified people and devices such as cameras, robots, smart phones, and drones as shown in Fig. \ref{smartcityarchitecture_4}.

\begin{figure*}[!h]
\centering
\includegraphics[width=5in]{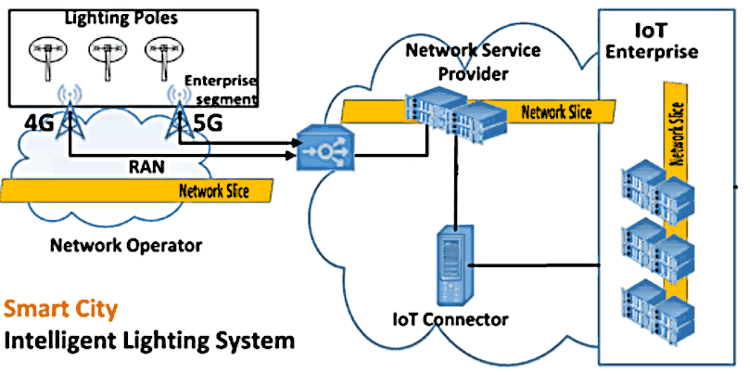}
\caption{Architecture of lighting management with 5G in smart cities \cite{rusti20195g}. In the architecture, 5G is used for data transmission for light poles. Similar to the above examples, 5G is treated as a simple component without details on the network in the whole architecture.}
\label{Architecture_of_lighting_management_with_5G}
\end{figure*}

The second subtype is that the studies connect a specific smart city scenario with 5G. The most mentioned scenario is lighting management (4 studies out of 30, 13.3\%). An example is shown in Fig. \ref{Architecture_of_lighting_management_with_5G}. In the scenario, a 5G small cell is deployed on a light pole with many other types of components (e.g., IoT sensors, advertisement screens, and surveillance cameras). User equipment and other components on the pole can connect with the 5G small cell for data transmitting and receiving. Besides using a 5G small cell, we could also use a 5G sensor on the pole connected to a 5G macrocell. In this way, components on the pole can use 5G services through the 5G sensor. Another example is the transportation scenario (3 studies out of 30, 10.0\%). With a 5G network using both macrocells and small cells that covers an area, vehicles (e.g., buses, autonomous cars, and trucks) can be connected in a 5G network. This can help them to share traffic information (e.g., traffic jams) and furthermore, to plan their routes or predict traffic events in a dynamic and real time way.

\subsubsection{5G network architecture in smart cities}
This type is related to a set of architectural design decisions of using 5G in smart cities. The decisions are regard different components in a 5G network, such as architecture of a macro or small cell (e.g., structure, band, height, range, modulation, and coding scheme), 5G deployment architecture (i.e., standalone, SA or non-standalone, NSA), cellular network architecture (e.g., C-RAN, coverage of the whole network, and number and distribution of base stations), and usage of 5G-related technologies (e.g., SDN and NFV). We provide three examples as follows. 

\begin{figure*}[!h]
\centering
\includegraphics[width=5in]{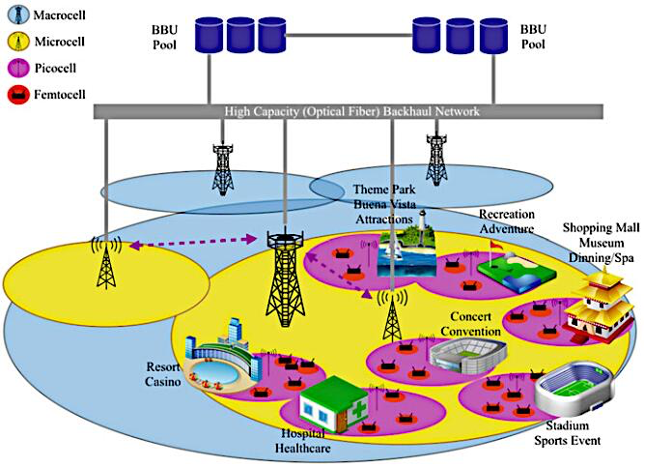}
\caption{High-level 5G network architecture in smart cities. The architecture includes both macrocells and small cells (i.e., microcells, picocells, and femtocells) with BBU pools \cite{liao20185g}.}
\label{High-level_5G_network_architecture_in_smart_cities}
\end{figure*}

\begin{figure*}[!h]
\centering
\includegraphics[width=5in]{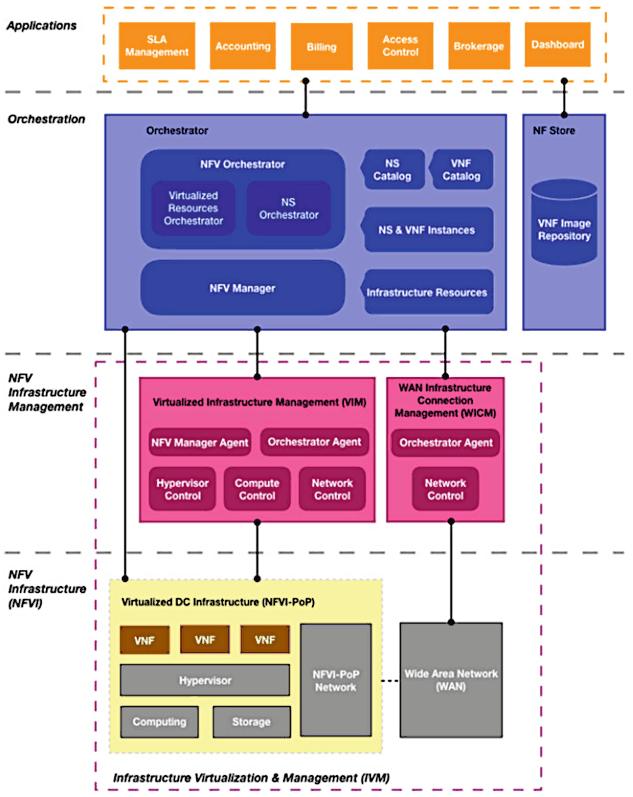}
\caption{System architecture of using 5G in smart cities \cite{liao20185g}. The architecture is based on the solution (e.g., SESAME) from the H2020 5GCity project. It has four layers: application layer, orchestration layer, NFVI infrastructure layer, and NFVI layer.}
\label{System_architecture_of_using_5G_in_smart_cities}
\end{figure*}

In \cite{liao20185g}, the authors focused on the design and applications of using 5G in casinos and entertainment in smart cities. Their high-level architecture includes both macrocells and small cells (i.e., microcells, picocells, and femtocells) with BBU pools to form their 5G network in a smart city as shown in Fig. \ref{High-level_5G_network_architecture_in_smart_cities}. In this architecture, antennas are deployed at macro and small cells, while BBUs are aggregated in one or several centralized locations (i.e., the C-RAN architecture). Moreover, different types of cells are used in different places. Macrocells are high-powered radio access nodes, which can cover a wide range. They are the basis of the 5G network. On the other side, small cells are the last-mile network infrastructure, which can complement the coverage from macrocells. For example, femtocells can only cover a small area and a few connections simultaneously, and therefore, femtocells can be used in smart homes in a building. Then the authors proposed a detailed system architecture of using 5G in smart cities based on the solution (e.g., SESAME) from the H2020 5GCity project as shown in Fig. \ref{System_architecture_of_using_5G_in_smart_cities}. The system architecture also has four layers as application layer (e.g., accounting, billing and access control), orchestration layer (e.g., NFV orchestrator and NFV manager), NFVI infrastructure layer (Virtualized Infrastructure Manager, VIM and wide area network infrastructure connection management), and NFVI layer (virtualized data center infrastructure and wide area network). The major concern is to enable multiple network operators to use the system through network slices in an edge-based and virtualized environment.

\begin{figure*}[!h]
\centering
\includegraphics[width=5in]{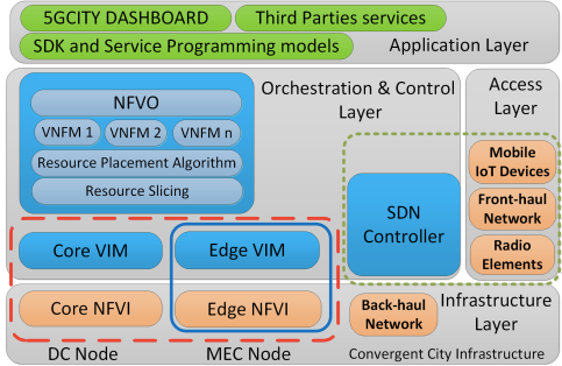}
\caption{An adapted 5G network architecture in smart cities \cite{paolino2019compute}. The architecture has four layers: application layer, orchestration and control layer, access layer, and infrastructure layer. Specifically, the architecture has multiple VIMs with edge VIMs, multiple NFVIs with edge NFVIs, and an SDN based wireless virtualization component.}
\label{An_adapted_5G_network_architecture_in_smart_cities}
\end{figure*}

In \cite{paolino2019compute}, the authors focused on virtualization of the 5G network, and proposed an adapted 5G network architecture in smart cities (see Fig. \ref{An_adapted_5G_network_architecture_in_smart_cities}). In the original architecture, it aims to deal with the integration of NFV, SDN, and edge computing based on a neutral host platform, in order to enable service providers to flexibly deploy applications and collaborate with each other, and also helps infrastructure owners to manage their investment. In such a neutral platform, virtualization is the key element, which provides abstraction of infrastructure resources. However, there could be various stakeholders with multiple heterogeneous applications interconnected in the platform. This leads to a problem regarding complicated virtualization that should support for example different stakeholders with multiple VIMs and orchestration solutions, as well as heterogeneous and distributed devices. To address the problem, their extended 5G network architecture has four layers, i.e., application layer (including third parties services, management dashboard, software development kit, and service programming models), orchestration and control layer (including NFV orchestrator, Virtualized Network Function (VNF) manager, VIM, resource management, core VIM, edge VIM, and SDN controller), access layer (including mobile IoT devices, front-haul network, and radio elements) and infrastructure layer (core NFV infrastructure, edge NFV infrastructure, and back-haul network). Such an architecture has multiple VIMs with edge VIMs, multiple NFVIs with edge NFVIs, and an SDN-based wireless virtualization component (for IoT devices, front-haul network, and radio elements). The idea of this architecture is to combine VIMs and deploy them in various point of presences and manage radio resources using an SDN controller for wireless virtualization. The architecture can also support different deployment scenarios.

In \cite{bronk2018technical}, the authors tried to address a practical problem, i.e., how 5G can be deployed in two Polish cities - TriCity and Warsaw. They first calculated the city-related data (e.g., number of devices and data traffic generated by the devices in the two cities) based on certain reports (e.g., \cite{cisco2012cisco}\cite{index2017zettabyte}). Then the authors further made a set of architectural design decisions and architectural assumptions regarding their 5G smart city network in their experiments. For decisions, they chose, for example, 700 MHz band, 46 dBM (40 W) transmitter power, 18 dBi antenna gains for base stations, QPSK, 16QAM, 64QAM, and 256QAM modulation, and four channel bandwidths as 5, 10, 15, and 20 MHz. For assumptions, examples are a. the number of residents in TriCity and Warsaw in 2021 will be 749,015 and 1,798,514 respectively; b. assuming the number of M2M devices in TriCity and Warsaw is 1.45 million and 3.48 million respectively; c. base station height is 30 m above terrain. The study clearly shows that different decisions and assumptions would have an impact on the performance and application of the 5G network in smart cities. For instance, based on all their architectural design decisions and architectural assumptions, they came up with that in a normal case of using only one 5MHz channel, TriCity needs 291 base stations and Warsaw needs 707 base stations to form a 5G network (see Table \ref{Number of base stations in TriCity and Warsaw based on different configurations}).

\begin{table*}[htbp]
\centering
\caption{Number of base stations in TriCity and Warsaw based on different configurations \cite{bronk2018technical}. The results are based on a set of design decisions (e.g., 700 MHz band and 46 dBM (40 W) transmitter power) and assumptions (e.g., the number of residents in TriCity and Warsaw in 2021 will be 749,015 and 1,798,514 respectively).}
\label{Number of base stations in TriCity and Warsaw based on different configurations}
\begin{tabular}{|p{0.13\columnwidth}|p{0.1\columnwidth}|p{0.1\columnwidth}|p{0.1\columnwidth}|p{0.1\columnwidth}|p{0.1\columnwidth}|p{0.1\columnwidth}|}
\hline
         & \multicolumn{2}{l|}{\textbf{TriCity}} & \multicolumn{2}{l|}{\textbf{Warsaw}} & \multicolumn{2}{p{0.2\columnwidth}|}{\textbf{Warsaw (worst case)}} \\ \hline
Channel bandwidth & Base station range (km) & Number of base stations & Base station range (km) & Number of base stations & Base station range (km) & Number of base stations \\ 
\hline
5 MHz    & 0.65               & 291              & 0.5               & 707              & 0.31                    & 1838                    \\ \hline
10 MHz   & 1.03               & 116              & 0.78              & 291              & 0.47                    & 800                     \\ \hline
15 MHz   & 1.27               & 77               & 0.98              & 184              & 0.6                     & 491                     \\ \hline
20 MHz   & 1.5                & 55               & 1.15              & 134              & 0.72                    & 341                     \\ \hline
2*15 MHz & 1.86               & 36               & 1.44              & 86               & 0.89                    & 223                     \\ \hline
2*20 MHz & 2.18               & 26               & 1.69              & 62               & 1.04                    & 164                     \\ \hline
3*15 MHz & 2.33               & 23               & 1.81              & 54               & 1.12                    & 141                     \\ \hline
3*20 MHz & 2.71               & 17               & 2.12              & 40               & 1.3                     & 105                     \\ \hline
\end{tabular}
\end{table*}

\subsubsection{Business architecture of using 5G in smart cities}
\begin{figure*}[!h]
\centering
\includegraphics[width=5in]{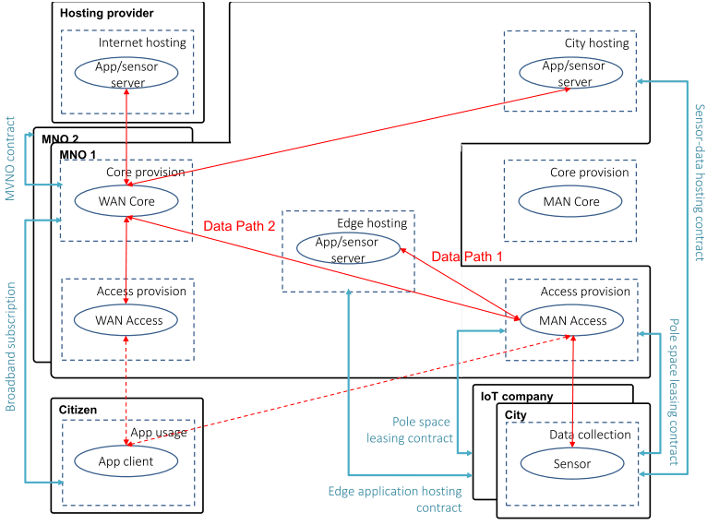}
\caption{Single MNO driven business architecture \cite{benseny2019city}. The context of the architecture is lighting management. The architecture shows the case of single MNO that only a single MNO has the rights to construct and exploit the 5G light pole network in smart cities.}
\label{Single_MNO_driven_business_architecture}
\end{figure*}

Not many studies talk the business architecture of using 5G in smart cities. However, we found that there are various types of stakeholders related to the usage of 5G in smart cities, such as government, company, service provider, operator, end user, and developer. Different business architecture used in a smart city may have various types of stakeholders with different roles, responsibilities, and concerns. In \cite{marabissi2019real}, the authors mentioned two business architecture, i.e., the open access architecture and the vertically integrated operator architecture. The open access architecture was proposed and deployed by Open Fiber in Prato. Network operators (i.e., who deployed the 5G network in Prato) are only infrastructure providers. They offer network slices to third parties such as virtual operators, companies, government, among others. These third parties then provide services to end users. The vertically integrated operator architecture was proposed and deployed by Wind-Tre in L’Aquila. Network operators play both the infrastructure provider role and the operator role who providing services to end users. In \cite{benseny2019city}, the authors discussed the same question regarding business roles in a 5G smart city. For example, in the analysis of the city network based on light poles, they identified four alternative decisions for business architecture: (1) single mobile network operator (MNO) driven business architecture (i.e., only a single MNO has the rights to construct and exploit the 5G light pole network), (2) joint MNO driven business architecture (i.e., multiple MNOs have the rights to construct and exploit the 5G light pole network), (3) city-driven business architecture (i.e., the city creates an operator company to construct and exploit the 5G light pole network), and (4) power company driven business architecture (i.e., only a single power company has the rights to construct and exploit the 5G light pole network). Different alternatives would bring different design of the business architecture. As an example, Fig. \ref{Single_MNO_driven_business_architecture} shows the single MNO driven business architecture.

\subsection{5G-related technologies used in/with smart city components}
This section reports the results of RQ3. We note that there are different classifications regarding whether a technology is smart city component or 5G component. Based on the 5G literature (e.g., \cite{agiwal2016next}\cite{gupta2015survey}\cite{yadav2018review}\cite{patil2012review}), in this study, we considered AI, IoT, big data, cloud computing, and block chain as smart city components instead of 5G-related technologies. The reason is that these technologies are usually not used in but with a 5G network in smart cities. For example, some studies use AI for object detection with a 5G network in a city transportation system. The most mentioned 5G-related technologies are radio access technologies (17 studies out of 30, 56.7\%), network slicing (13 studies out of 30, 43.3\%), and edge computing (10 studies out of 30, 33.3\%). These technologies can also be interconnected with each other in a 5G network. Details of the three types of 5G-related technologies are provided as follows.

\subsubsection{Radio access}
Radio access technologies include a set of different focuses, for example, resource control, base station grouping, energy consumption scheme, small cells and antenna design. We provide three examples as follows. In \cite{han20175g}, the authors proposed a cell-less 5G network. In the network, they used a novel radio access technology: for uplinks, devices broadcast and nearby base stations receive the data and forward the data to the SDN cloud for processing (e.g., decoding); for downlinks, the SDN cloud makes a decision regarding which one or several base stations should transmit the data to the devices based on location and channel status of the devices. In \cite{habbal2019context}, the authors proposed a context-aware radio access technology. The idea is that a 5G network would be heterogeneous, i.e., including different radio access technologies, and therefore, the 5G network should be aware of the context of each radio access technology (e.g., suitability) as well as user equipment (e.g., movement). Context management in the studies includes context sensing (i.e., gather context information from both devices and network), context aggregation (i.e., aggregating data from multiple dimensions), context processing (processing the aggregated data to form a context based on a set of criteria), and context storing (i.e., storing context information in a repository). In \cite{sharif2019compact}, the authors proposed an antenna design technology for base stations in a parking scenario of smart cities. The antenna has enhanced directivity and narrower beamwidth, providing a bandwidth ranging from 6GHz to 7.25 GHz with 7 dB gain and 110 degrees of half-power beamwidth. In the parking scenario of smart cities, the antenna can be used to improve the real-time location accuracy, which is especially useful for autonomous cars’ parking.

\subsubsection{Network slicing}
5G network slicing divides a physical 5G network infrastructure into various virtualized and independent logical units. Each slice is an end to end network serving for a specific scenario, running by one or more virtual network operators. NFV and SDN are the most critical technologies to realize network slicing. We provide three examples as follows. In \cite{marabissi2019real}, the authors developed a 5G system, composed of three layers, i.e., physical infrastructure layer, network function layer, and service layer. In the second layer, the system has two planes, i.e., user plane and control plane. Network functions are configured both in the user plane and control plane for each slice. Then based on the configuration of each slide, the third layer provide different services. They also defined key performance indicators for the planes (e.g., time to switch from an idle or inactive state to an active state should be less than 20 ms for the uRLLC and eMBB scenario). In \cite{storck20195g}, the authors designed a 5G core, which used SDN and virtualization technologies. The design separated the data plane from the control plane. Through using NFV and SDN, the 5G network can provide specific virtual networks with different services based on the same physical infrastructure. In \cite{wu2019dynamic}, the authors also presented a network slicing based 5G core network connecting with heterogeneous networks. They specifically used Multiprotocol Label Switching (MPLS) in network slicing. Therefore, the core network has customer edge routers (i.e., connecting with heterogeneous networks), provider edge routers (acting as interfaces between heterogeneous networks and optical core), and provider routers (forwarding data traffic from customer edge routers and provider edge routers).

\subsubsection{Edge computing}
Edge computing is also an important technology identified from the studies. The terms could be various (e.g., mobile edge computing and multi-access edge computing). Though many studies claim that edge computing is part of their architecture, we did not get much details of this technology, i.e., how exactly edge computing was used in the 5G network in smart cities. We provide three examples as follows. In \cite{liao20185g}, the idea is to deploy applications closer to 5G users, in order to reduce data congestion of the 5G network and provide better services. The system used available resources at the RAN side to deploy the edge computing components. In \cite{paolino2019compute}, the authors proposed a novel component named edge VIM and edge NFVI. A specific characteristic of the proposed edge VIM and edge NFVI is that they can be used to support trust computing for devices based on an ARM architecture. Through providing a virtualization-based security and trust infrastructure, the system supports functions such as authentication, asset tagging, and secure storage. In \cite{raj2018enabling}, the authors used edge computing to address data traffic problems and reduce end-to-end latency in the smart mall scenario of smart cities. They also merged edge computing with the NFV technology, which can provide high bandwidth and low latency, enhance resource usage, and deploy automatic services.

\subsection{Challenges (problems) of using 5G in smart cities}
This section reports the results of RQ4. We identified four types of challenges of using 5G in smart cities: complex context, challenging requirements, 5G network development, and miscellaneous.

\subsubsection{Complex context of using 5G in smart cities}
The context of using 5G in smart cities are complex, which brings many challenges. We provide three examples as follows. In \cite{lynggaard2015deploying}, the authors stated that 5G would not be alone in smart cities, but more like a component in heterogeneous networks, cooperated working with other components (e.g., IoT, Wifi, and 2G/3G/4G). How the heterogeneous networks (including 5G) work in smart cities is a challenge. Moreover, there would be many devices deployed everywhere in a smart city. These devices can generate massive data that needs to be stored and processed through the heterogeneous networks. Finally, there would be, for example, many walls in a smart city, and penetration of walls is a challenge when using 5G. In \cite{vo20185g}, the authors mentioned that the integration of wireless sensor networks (WSNs) into the 5G network in smart cities is difficult. One major reason is because of the nature of WSNs, i.e., vast diversity. Nodes in WSNs can be extremely heterogeneous. Energy cost for transmission and data delivery capacity performance are other concerns. They pointed out the problem regarding huge number of devices and the generated data make the network congested. In \cite{wu2019dynamic}, the authors mentioned that data traffic in heterogeneous networks of smart cities is dynamic, because the interaction among heterogeneous networks could fluctuate for many reasons. Therefore, managing resources efficiently in such a dynamic context is challenging.

\subsubsection{Challenging requirements of using 5G in smart cities}
Though 5G is designed to meet a set of requirements, some of them are challenging. We provide three examples as follows. In \cite{condoluci2015softwarization}, the authors discussed the concerns of security of using 5G in smart cities. A 5G network can include various types of stakeholders (e.g., network operators and third parties) based on different business modes. When using technologies such as network slicing, data are stored and shared through the same physical infrastructure. Therefore, it needs careful thinking regarding security in such a system. In \cite{paolino2019compute}, though the authors agreed that security is important in smart cities and personal data needs to be well protected, they also mentioned that the added benefits of security come at the price of lower the performance of the system. For many requirements of using 5G in smart cities, usually there is a balance to make based on pros and cons. In \cite{qi2019uav}, the authors designed a UAV-based 5G network for smart cities. They used a leader UAV and a set of normal UAVs with base stations to form a hierarchical 5G network including the area of the sky. In this specific context, there is a challenge regarding the battery of the UAVs. Without a reasonable battery life, the UAV-based system will not work properly.

\subsubsection{5G network development in smart cities}
Considering the complex context and challenging requirements of using 5G in smart cities, 5G network development is difficult, such as design and implementation of 5G-related technologies, including network slicing, radio access, deployment scheme, architecture, interference, and handover. We provide three examples as follows. In \cite{storck20195g}, the authors analyzed the design of a 5G project named Novel Radio Multiservice Adaptive project (NORMA). They mentioned that some infrastructure in the network may contain components that used by multiple network slices and do not support virtualization. Even for the components that support virtualization, the virtualization technology used is far from maturity to allow NFV. In \cite{marabissi2019real}, the authors claimed that efficiently managing network slices is challenging. This includes efficient resource allocation, virtualizing and dividing the RAN into different slices, among others. In \cite{condoluci2015softwarization}, the authors mentioned that it is unclear how existing SDN controllers of a 5G network perform in a wide area (e.g., a smart city). In other words, with a huge number of devices, load of data, and simultaneously connections in a smart city, how to use 5G-related technologies (e.g., SDN) to deal with, for example, scalability of the 5G network, 5G network overloading, and 5G network congestion, is a critical challenge.

\subsubsection{Miscellaneous}
We also found some other types of challenges regarding using 5G in smart cities. Examples are lack of tools, cooperation between 5G and other technologies (e.g., AI, IoT, and cloud computing), uncertainties with assumptions, and poor infrastructure. We provide three examples as follows. In \cite{condoluci2015softwarization}, the authors stated that though IoT devices are one major type of smart city components, using 5G for the communications of IoT devices could be challenging. For instance, 5G should be able to deal with the transmission of limited data (e.g., few bytes) in an efficient way, requiring new protocol interfaces in the SDN/NFV based architecture. In \cite{rusti2018smart}, the authors pointed out that the main challenge is the lack of networking and computational infrastructure to support industry vertical requirements through using network slicing. In \cite{gholampooryazdi2017scenario}, the authors mentioned that there are many uncertainties regarding for example technologies, applications, services, and stakeholders related to having 5G smart light poles in cities. An example of such uncertainties is about who owns and operates the platforms and services of 5G smart light poles. The uncertainties drive many assumptions, which can challenge the decision making process of using 5G in smart cities.

\subsection{Lessons learned from using 5G in smart cities}
This section reports the results of RQ5. For the lessons learned, we note that they are regarding the experience of 5G usage in smart cities based on evaluations, instead of experience about how to conduct evaluations. 22 studies (out of 32, 68.8\%) report benefits regarding 5G itself or their proposed 5G-related methods in smart cities. We provide three examples as follows. In \cite{Okumura}, the authors confirmed that in the public safety scenario with surveillance cameras, using 5G can increase video frame rate, in order to increase face recognition frequency up to six times, compared to using 4G. In \cite{han20175g}, the authors proposed a 5G converged cell-less communication method for smart cities. They claimed that based on the simulation results, their method can improve the coverage probability and energy saving at base stations as well as mobile devices, which can fulfill the requirements of coverage and data rate in smart cities. In \cite{oproiu20185g}, the authors mentioned that based on their design for the 5G lighting management scenario in smart cities, it addresses three major functions that the existing lighting system is not able (or difficult) to fulfill: real-time and remote control with high security of light poles, efficient energy management, and proactively malfunction identification.

14 studies (out of 32, 43.8\%) present lessons learned regarding 5G-related technologies or requirements on the technologies. We provide three examples as follows. In \cite{qi2019uav}, the authors proposed a method including a leader UAV and a set of normal UAVs with base stations to form a hierarchical 5G network including the area of the sky for smart cities. The authors stated that if the flight trajectory of the leader UAV can be analyzed by base stations, the beamforming can be more precise, in order to improve data rate of the UAV-based 5G network. In \cite{bronk2018technical}, the authors concluded that there are different configurations of using 700 MHz 5G network that can work in both TriCity and Warsaw. An extreme example could be to use three aggregated 20 MHz channels. In this circumstance, TriCity needs 17 base stations and Warsaw needs 40 base stations. In \cite{oproiu2017development}, the authors concluded that a 5G smart city lighting system should fulfill the following requirements: low bandwidth needs, low delay, fast deployment, large number of devices and sensors with massive communications, among others.

There are other types of lessons learned (8 studies out of 32, 25.0\%) regarding for example cost, business, stakeholders, and cooperation of using 5G in smart cities. We provide three examples as follows. In \cite{karadimce2018smart}, the authors mentioned that the future smart cities depend not only on the 5G infrastructure, but also the availability and quality of information sharing and social interaction. In \cite{rao2018impact}, the authors considered that 5G would change the exiting value chains and enable new opportunities (e.g., new applications and employment) for smart cities. This would further lead to enhanced economic growth (e.g., growth of GDP). In \cite{schneir2019business}, the authors calculated the cost of deploying a 5G network in the Central London area. The results show that the return on investment is 29\% and the net present value is 35 million.

\section {Discussion}\label{section5}
This section presents the analysis of the results and the implications for researchers and practitioners.
\subsection{Analysis of the results}
According to several surveys and reviews on the 5G topic (e.g., \cite{agiwal2016next}\cite{gupta2015survey}\cite{yadav2018review}\cite{patil2012review}), there are many studies concerning different aspects of 5G (e.g., 5G technologies and applications). However, much less studies (i.e., 32 studies in this SMS) talk about using 5G in smart cities, and less than 5 studies were published before year 2017. Moreover, most of the studies identified in this SMS only demonstrate using 5G in smart cities with toy examples or academic studies (e.g., simulations). There are three potential reasons: (1) Though the research of 5G started from 2012, using 5G in smart cities is still on its early stage, and cities only launched certain number of 5G trial projects. (2) Using 5G in smart cities is complex, including different scenarios and concerns, leading to various system design and implementation. The scenarios and concerns have not been comprehensively extracted and analyzed. (3) Research or applications of using 5G in smart cities require various types of resources (e.g., policy, standards, stakeholders, and physical or virtual infrastructure). Coordinating these resources could be a large amount of endeavor. Finally, 16 studies (out of 32, 50.0\%) include assumptions regarding 5G network design or smart city scenarios. Though certain assumptions in the studies are based on other published literature or the authors claimed that they were confident about the validation of the assumptions they made, there is always a risk when the assumptions turn to be invalid in certain context. Invalid assumptions can lead to the evidence from the studies being invalid, and further make the reported design and implementation of the 5G network be problematic in applications.

We did not find any systematic classification of the scenarios regarding using 5G in smart cities from the studies. The main reason is that smart cities are complex systems and include various types of components. Instead of looking at the whole picture, the selected studies focus on a small part of smart cities (e.g., transportation). Moreover, it is interesting that lighting management received much attention than the others. Though there would be many light poles in a city and it is important to make these poles be connected through 5G in the future, lighting management is a relatively small area in a smart city. One major reason is that the 5G Infrastructure Public Private Partnership (5G PPP) plays an important role in prompting the usage of the 5G-based lighting management in smart cities. Two projects from the 5G PPP (i.e., MATILDA and SLICENET) contain the lighting management scenario, and 4 selected studies (out of 9, 44.4\%) were supported by the two projects. Besides, another project named LuxTurrim5G (a lighting management project) also supported 3 selected studies (out of 9, 33.3\%). The results also show that 5G is related to many types of components in smart cities. These can also bring new business opportunities and patterns. As an example, 5G is only an ICT, which should work with other technologies (e.g., AI, IoT, big data, and cloud computing) in smart city scenarios. As another example, constructing, operating, and maintaining a 5G network are related to various types of stakeholders and may need to be adapted according to the policies or strategies employed in different cities. The identified components are different from the smart city components mentioned in the existing literature (e.g., \cite{arroub2016literature} and \cite{chamoso2018tendencies}). The reasons are that: (1) The SMS and the existing literature have different scope (i.e., we focused on only using 5G in smart cities); (2) we used the Constant Comparison method to analyze the selected studies, and the component types were generated during the comparison process since we did not predefine any type of the components.

There are three types of architecture proposed and/or discussed by the studies, i.e., smart city architecture (treating 5G as a component), 5G network architecture in smart cities, and business architecture of using 5G in smart cities. For the first type, the differences among the architecture are more regarding the specific smart city scenarios. In other words, 5G is considered as an ICT component used in one or more smart city scenarios for communication purposes. Details of the 5G network itself are usually abstract or unknown. For the second type, the results are consistent with other 5G-related studies. For example, there are discussions in the studies regarding using CU and DU instead of the traditional BBU. Also we found many other architectural design decisions (e.g., whether to use the C-RAN architecture, which band to choose, and the height of a macro base station) that have to be made when designing and implementing a 5G network in smart cities. Though we believe that managing such architectural design decisions are important for the success of using 5G in smart cities, the decisions are usually scattered in the studies and have not been systematically studied. Finally, not many studies talk about the business architecture, but the studies mention various types of stakeholders of using 5G in smart cities. The reason is that when proposing a method in the context of using 5G in smart cities, stakeholders are usually an important part of the method. Another reason is that 5G is still on its early stage, and therefore, there are many uncertainties regarding the business component.

Radio access, network slicing, and edge computing are the most discussed 5G-related technologies in smart city scenarios. This is consistent with the results in existing 5G-related studies. However, most of the studies lack details regarding, for example, how to cut a 5G network into slices or how to virtualize 5G network components. One reason is that such technologies are assumed to be common knowledge. Moreover, we did not get much data regarding the physical layer and MAC layer of a 5G network (e.g., mm-wave, beamforming, antenna, MAC protocols, modulation, and coding) from the studies. The main reason is that the topic of the SMS is about using 5G in smart cities instead of 5G itself. The focus of the studies is more related to the application level.

We identified four types of challenges of using 5G in smart cities: complex context, challenging requirements, 5G network development, and miscellaneous. The major reason is that 5G is an ongoing technology. Especially regarding the usage of 5G in smart cities, there are many related scenarios, components (e.g., technologies, infrastructure, policy, and business patterns), and uncertainties that need to be considered. We believe that it needs more time to think, design, construct, and maintain 5G networks in smart cities.

Most of the identified lessons learned are the benefits of 5G itself or the proposed 5G-related methods in smart cities. There is a lack of lessons learned from industrial experience on this topic. The main reason is that the majority of the studies (30 studies out of 32, 93.8\%) only have demonstration, provide toy examples, or use academic studies, instead of having industrial studies or industrial practice.

\subsection{Implications for researchers and practitioners}
The implications for researchers and practitioners are summarized as follows.

(1) On the one hand, smart city is one of the most important applications in the 5G era. On the other hand, there are only 32 studies identified and selected from the time period between January 2012 and December 2019 on using 5G in smart cities. Most of the studies only focus on a small part of this topic and do not have industrial evidence (i.e., Level 5 or 6). Researchers are encouraged to design and conduct more empirical studies (e.g., case studies, experiments, and surveys) with practitioners on the topic of using 5G in smart cities.

(2) Implicit and invalid assumptions may cause problems. Researchers are encouraged to make all their assumptions (or at least critical ones) explicit and consider more carefully whether their assumptions would turn to be invalid in certain circumstances.

(3) Not every scenario of using 5G in smart cities received the same attention from the studies. We also did not see a systematic classification of scenarios. It is valuable to first identify what scenarios are when using 5G in smart cities and then analyze for example their impacts and importance. This will help researchers to get a better understanding of the topic. Moreover, the mapping from a smart city scenario to a 5G pattern (i.e., eMBB, mMTC, and uRLLC) is not always clear mentioned in the studies. We suggest that researchers should make this mapping more explicit. 

(4) Though we identified six types of smart city components related to 5G, the relationships between scenarios, smart city components, and 5G architecture and technologies in the 5G-enabled smart city system are not well studied, which needs further investigation.

(5) The first type of the identified architecture (i.e., smart city architecture) is more like background (or concept) with limited information. Moreover, certain 5G technologies (e.g., edge computing) are usually presented without enough details (i.e., assuming that they are common knowledge). This may impede researchers and practitioners to reproduce the results of the study or take the proposed architecture and technologies in practice.

(6) Lessons learned are important to help researchers and practitioners to successfully use 5G in smart cities. Not only positive results are needed, negative results could be more valuable since they can help to avoid potential problems. Researchers and practitioners are encouraged to share their experience on both positive and negative side of using 5G in smart cities.

(7) Though some cities started 5G trial projects, overall it is far from mature. At the current stage, with the consideration of incomplete city infrastructure, immature 5G-related technologies, complex context, among others, using 5G in smart cities has many unknowns and needs a long way to evolve. Practitioners are encouraged to be actively involved in the 5G development and evaluation processes.

(8) 5G is never separate in smart cities, but it is closely related to many other types of smart city components (e.g., stakeholders, city infrastructure, and business patterns). When deploying a 5G network in a smart city scenario, practitioners need to identify the related components as well as their relationships and impacts, in order to make reasonable decisions.

(9) Though there are certain findings (e.g., the proposed methods) of using 5G in smart cities, they do not come for free, but are usually constrained by specific context (e.g., with assumptions). Before taking any existing finding to practice, practitioners need to carefully consider whether the findings match their specific context.

\section {Conclusions}\label{section6}
In this work, we conducted an SMS that covers literature published between January 2012 and December 2019 on using 5G in smart cities. We conclude the SMS as follows.
(1) We finally got 32 selected studies, which are distributed over 27 publication venues, containing 17 conferences (out of 27, 63.0\%), 9 journals (out of 27, 33.3\%), and 1 book (out of 27, 3.7\%). 17 studies (out of 32, 53.1\%) report results based on academic studies and 13 studies (out of 32, 40.6\%) use demonstration or toy examples. Only 2 studies (out of 32, 6.3\%) report using 5G in smart cities based on industrial studies. 16 studies (out of 32, 50.0\%) include assumptions of 5G network design or smart city scenarios.
(2) The most discussed smart city scenario is transportation, followed by public safety, healthcare, city tourism, entertainment, and education.
(3) 28 studies (out of 32, 87.5\%) propose and/or discuss the architecture of 5G-enabled smart cities, containing smart city architecture (treating 5G as a component), 5G network architecture in smart cities, and business architecture of using 5G in smart cities.
(4) The most mentioned 5G-related technologies are radio access technologies (17 studies out of 30, 56.7\%), network slicing (13 studies out of 30, 43.3\%), and edge computing (10 studies out of 30, 33.3\%). 
(5) The challenges are mainly about complex context, challenging requirements, and network development of using 5G in smart cities.
(6) Most of the lessons learned identified are benefits regarding 5G itself or the proposed 5G-related methods in smart cities.

This is the first systematic analysis that specifically focuses on the topic of using 5G in smart cities. 
Most of the studies identified in this SMS only demonstrate using 5G in smart cities with toy examples or academic studies (e.g., simulations). The context of using 5G in smart cities is rather complex and certain requirements are difficult to fulfill, leading to that the development of 5G in smart cities is usually case by case (i.e., lacking of generalization) and there are many technical, management, and social uncertainties and problems emerging in practice. Moreover, though there are various policies by different governments and white papers by different companies or other organizations in the field of 5G, they are usually abstract and not specific for smart cities, i.e., we cannot actually use them to design and implement the 5G network in a smart city. We still need more industrial studies of using 5G in smart cities, in order to collect more data from different cases, and further analyze the architecture, design, and even business patterns from the data.
Besides the descriptions and discussions presented in Section 4 and Section 5, there are many other interesting areas of using 5G, e.g., D2D communications, virtual reality, augmented reality, smart industry, and smart grid. These areas are also under development, which need further research.

\appendix

\section{Terms and their abbreviations}
AAU: Active Antenna Unit

AI: Artificial Intelligence

BBU: Baseband Unit

C-RAN: Cloud Radio Access Network

CU: Centralized Unit

D-RAN: Distributed Radio Access Network

DU: Distributed Unit

D2D: Device to Device

eMBB: enhanced Mobile Broadband

ICTs: Information and Communication Technologies

IoT: Internet of Things

MIMO: Multiple Input Multiple Output

mm-wave: millimeter wave

mMTC: massive Machine Type Communications

MNO: Mobile Network Operator

MPLS: Multiprotocol Label Switching

M2M: Machine to Machine

NFV: Network Functions Virtualization

NSA: Non-standalone

OFDM: Orthogonal Frequency Division Multiplexing

RQ: Research Question

RRH: Remote Radio Head

RRU: Remote Radio Unit

SA: Standalone

SDN: Software Defined Network

SLR: Systematic Literature Review

SMS: Systematic Mapping Study

UAV: Unmanned Aerial Vehicle

uRLLC: ultra-Reliable and Low-Latency Communications

VIM: Virtualized Infrastructure Manager

VNF: Virtualized Network Function

V2I: Vehicle to Infrastructure Communication

V2P: Vehicle to Pedestrian Communication

V2V: Vehicle to Vehicle Communication

V2X: Vehicle to Everything

WSN: Wireless Sensor Network

\section{Study results}

\begin{table*}[!htbp]
\centering
\caption{Sources of the studies on using 5G in smart cities. The 32 studies are published in 27 publication venues, containing 17 conferences (out of 27, 63.0\%), 9 journals (out of 27, 33.3\%), and 1 book (out of 27, 3.7\%).}
\label{Sources of the studies on using 5G in smart cities}
\begin{tabular}{|p{0.6\columnwidth}|p{0.15\columnwidth}|p{0.15\columnwidth}|}
\hline
\textbf{Venue}                                                              & \textbf{Type} & \textbf{Number (\%)} \\ \hline
IEEE Communications Magazine                                                & Journal       & 2 (6.3\%)            \\ \hline
Wireless Personal Communications                                            & Journal       & 2 (6.3\%)            \\ \hline
IEEE Access                                                                 & Journal       & 2 (6.3\%)            \\ \hline
Smart Cities \& Information and Communication Technology (CTTE-FITCE) & Conference & 2 (6.3\%) \\ 
\hline
International Conference on Communications (COMM)                           & Conference    & 2 (6.3\%)            \\ \hline
Wireless Communications and Mobile Computing                                & Journal       & 1 (3.1\%)            \\ \hline
NTT Technical Review                                                        & Journal       & 1 (3.1\%)            \\ \hline
IEEE Network                                                                & Journal       & 1 (3.1\%)            \\ \hline
Journal of Network and Computer Applications                                & Journal       & 1 (3.1\%)            \\ \hline
Future Internet                                                             & Journal       & 1 (3.1\%)            \\ \hline
Telecommunications Policy                                                   & Journal       & 1 (3.1\%)            \\ \hline
IEEE 5G World Forum (5GWF)                                                  & Conference    & 1 (3.1\%)            \\ \hline
International Conference on Developments in Systems Engineering (DeSE)    & Conference    & 1 (3.1\%)            \\ \hline
Telecommunication Forum (TELFOR)                                            & Conference    & 1 (3.1\%)            \\ \hline
International Conference on Innovations for Community Services (I4CS)     & Conference    & 1 (3.1\%)            \\ \hline
IFIP/IEEE Symposium on Integrated Network Management (IM)                   & Conference    & 1 (3.1\%)            \\ \hline
International Conference on Computational Intelligence and Computing Research (ICCIC) & Conference & 1 (3.1\%) \\ \hline
AEIT International Annual Conference                                        & Conference    & 1 (3.1\%)            \\ \hline
Joint CTTE and CMI Conference on Internet of Things – Business Models, Users, and Networks & Conference  & 1 (3.1\%) \\ \hline
Engineering Software Systems: Research and Praxis                           & Conference    & 1 (3.1\%)            \\ \hline
International Conference on Contemporary Computing and Informatics   (IC3I) & Conference    & 1 (3.1\%)            \\ \hline
International Congress on Ultra Modern Telecommunications and Control Systems and Workshops (ICUMT) & Conference & 1 (3.1\%) \\ \hline
International Internet of Things Summit                                     & Conference    & 1 (3.1\%)            \\ \hline
Baltic URSI Symposium (URSI)                                                & Conference    & 1 (3.1\%)            \\ \hline
IEEE Latin-American Conference on Communications (LATINCOM)                 & Conference    & 1 (3.1\%)            \\ \hline
International Conference on Computational Science and Computational Intelligence (CSCI) & Conference & 1 (3.1\%) \\ \hline
Handbook of Smart Cities                                                    & Book          & 1 (3.1\%)            \\ \hline
\end{tabular}
\end{table*}

\begin{figure}[!htbp]
\centering
\includegraphics[width=2.8in,height=10cm]{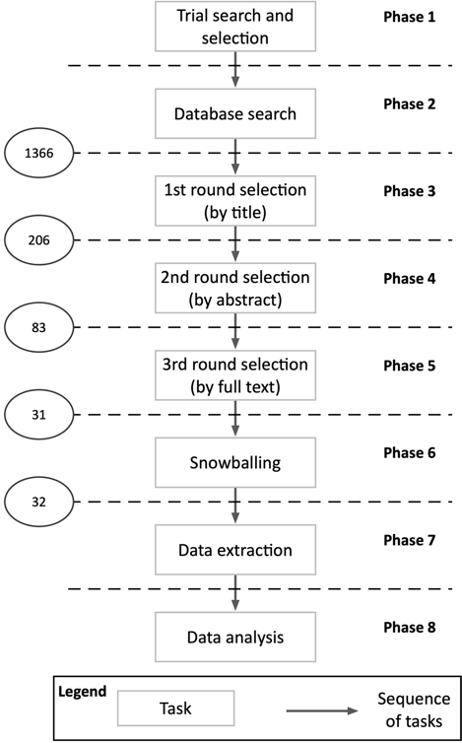}
\caption{Search and selection results of the studies. The numbers in the figure denote the number of studies from each phase of the study search and selection process.}
\label{Search_and_selection_results_of_the_studies}
\end{figure}

\section*{Acknowledgments}
This work has been partially supported by the National Key R\&D Program of China with Grant No. 2018YFB1402800 and Research Foundation of Shenzhen Polytechnic.

\makeatletter
\renewcommand{\@biblabel}[1]{[#1]}
\makeatother
\renewcommand{\refname}{References}
\bibliographystyle{IEEEtran}
\bibliography{ref.bib}
\balance
\end{document}